\documentclass[pra,aps,twocolumn,nofootinbib,showpacs,floatfix,superscriptaddress,amsmath]{revtex4}
\usepackage{graphicx}
\usepackage{amsmath}
\usepackage{amssymb}

\newcommand {\fexp} [1] {\exp \left( #1 \right)}

\newcommand {\fsin} [1] {\sin \left( #1 \right)}
\newcommand {\fabsq}[1] {\left| #1 \right|^2}
\newcommand {\la}{\langle}
\newcommand {\ra}{\rangle}
\begin{document}
\title{Inhibiting unwanted transitions in population
  transfer in two- and three-level quantum systems}
\author{A. Kiely}
\email{anthony.kiely@umail.ucc.ie}
\affiliation{Department of Physics, University College Cork, Cork, Ireland}
\author{A. Ruschhaupt}
\email{aruschhaupt@ucc.ie}
\affiliation{Department of Physics, University College Cork, Cork, Ireland}
\begin{abstract}
We examine the stability of population transfer in two- and three-level systems against unwanted additional transitions. This population inversion is achieved by using recently proposed schemes called ``shortcuts to adiabaticity''.
We quantify and compare the sensitivity of different schemes to these unwanted transitions. Finally, we provide examples of shortcut schemes which lead to a zero transition sensitivity in certain regimes, i.e. which lead to a nearly perfect population inversion even in the presence of unwanted transitions.
\end{abstract}
\maketitle


\section{Introduction}

The manipulation of the state of a quantum system with time-dependent interacting
fields is a fundamental operation in atomic and molecular physics.
Modern applications of this quantum control such as quantum information
processing \cite{NC} require fast schemes with a high fidelity (typically with
an error lower than $10^{-4}$ \cite{NC}) which must also be very stable with
respect to imperfections of the system or fluctuations of the control
parameters.

Most methods used may be classified into two major groups:  
fast, resonant, fixed-area pulses, and slow  
adiabatic methods such as ``Rapid'' Adiabatic Passage (RAP).  
Fixed area pulses are traditionally considered to be fast but unstable with
respect to perturbations. For two-level systems, an example of a fixed area pulse is a $\pi$ pulse.
A $\pi$ pulse may be fast  but is highly sensitive to variations in the pulse
area and to inhomogeneities in the sample \cite{Allen}. An alternative to a
single $\pi$ pulse are composite pulses \cite{Levitt,Collin,Torosov}, 
which still need an accurate control of pulse phase and intensity.
On the other hand, the canonical robust option is to perform operations adiabatically \cite{adiab}.
Nevertheless, such schemes are slow and therefore likely to be affected by decoherence or noise  
over the long times required and do not lead to an exact transfer.

A compromise is to use  
``shortcuts to adiabaticity'' (STA), which may be broadly defined as the   
processes that lead to the same final populations as the adiabatic
approach but in a shorter time, for a review see \cite{sta_review, sta_review_2}. 
In particular, STA for two- and three-level systems are
developed in \cite{Chen2010, Sara12, noise, Andreas2013,Sara13} and \cite{sta_3level} respectively. 

Nonetheless, in an experimental implementation, the system is never
an ideal two- or three-level system. There may be unwanted couplings to
other levels. The effect of such unwanted transitions for composite pulses has
been examined and optimized in \cite{genov_2013} where it was also assumed
that the phase of the unwanted coupling to another level could be controlled in a time-dependent way.
The goal of this paper is  to examine the effect of unwanted couplings to
STA in two- and three-level quantum systems where we will assume that no control of the phase of the coupling to the unwanted level is possible. 

The remainder of this paper is structured as follows. In the subsequent section, we briefly
review STA for two-level systems. In Section \ref{sect3},
we examine the sensitivity of STA schemes to unwanted transitions and present
schemes to minimize this sensitivity. 
In Section \ref{sect4}, we review STA for three-level systems and we will examine and
optimize their sensitivity to unwanted transitions in Section \ref{sect5}.


\section{Invariant-based shortcuts in two-level quantum systems\label{sect2}}

\begin{figure}[t]
\begin{center}
(a) \includegraphics[angle=0,width=0.3\linewidth]{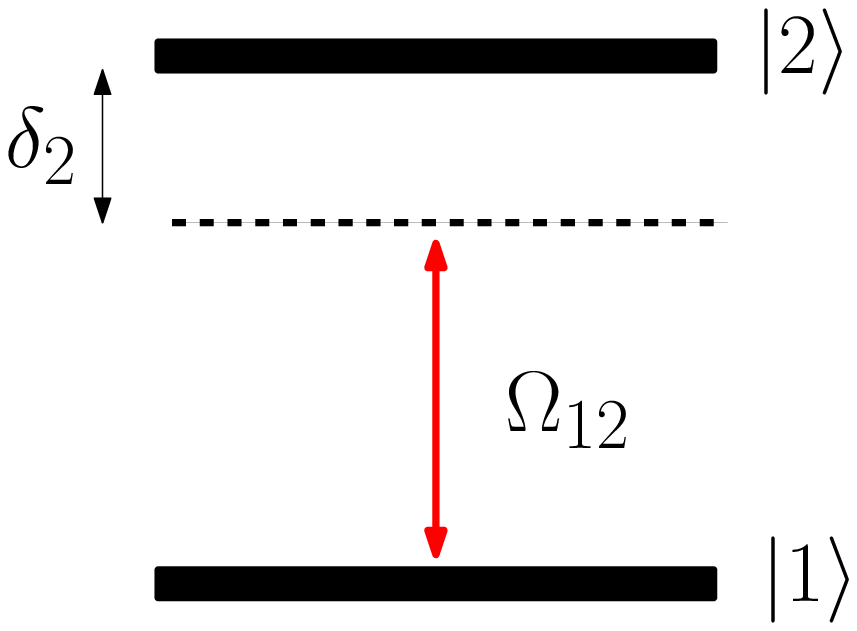}\\[0.5cm]

(b) \includegraphics[angle=0,width=0.6\linewidth]{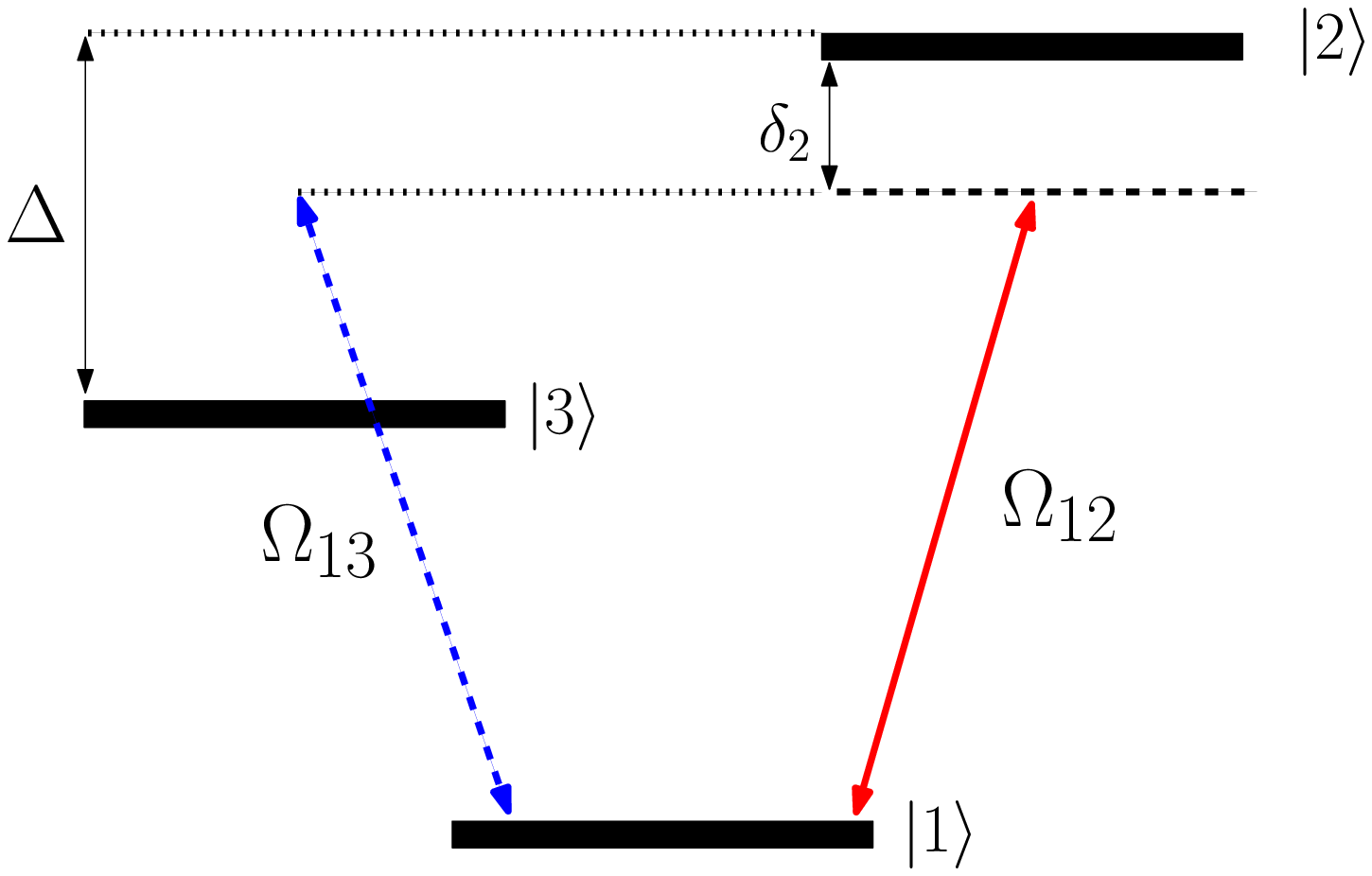}\\[0.5cm]

(c) \includegraphics[angle=0,width=0.6\linewidth]{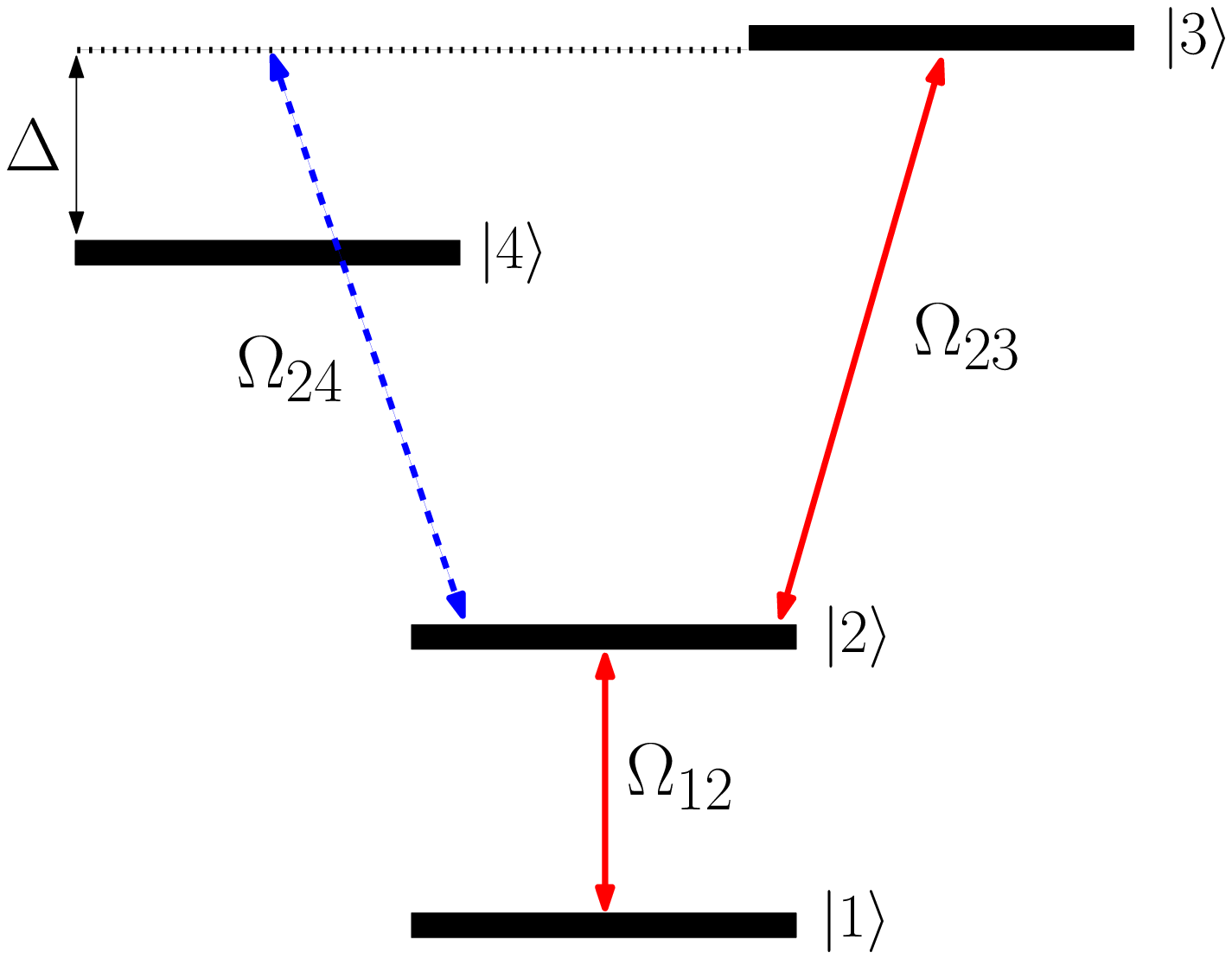}
\end{center}
\caption{\label{fig_1_basics} (Color online) Schematic of level structure: (a) Ideal
  two-level system; (b) two-level system with an unwanted coupling (blue dotted arrow) to a third level;
(c) three-level system with an unwanted coupling (blue dotted arrow) to a fourth level.}  
\end{figure}

Here we will review the derivation of invariant-based STA schemes in two-level quantum systems
following the explanation given in \cite{noise}.
We assume our two-level system has a Hamiltonian of the form
\begin{equation}
H_{2L}(t)=\frac{\hbar}{2}\left(\begin{array}{cc}
-\delta_{2}(t) & \Omega_{R}(t)-i\Omega_{I}(t)\\
\Omega_{R}(t)+i\Omega_{I}(t) & \delta_{2}(t)
\end{array}\right)\label{eq:1}
\end{equation}
where the ground state is represented by 
$\left|1\right\rangle =\left(\begin{array}{c}
1\\
0
\end{array}\right)$
and the excited state by $\left|2\right\rangle =\left(\begin{array}{c}
0\\
1
\end{array}\right)$ as in Fig. \ref{fig_1_basics}(a).

An example of such a quantum system would be a semiclassical coupling of two atomic levels with a laser
in a laser-adapted interaction picture. In that setting $\Omega(t) = \Omega_R(t) + i \Omega_I(t)$ would be the complex Rabi
frequency (where $\Omega_R$ and $\Omega_I$ are the real and imaginary parts)
and $\delta_{2}$ would be the time-dependent detuning between transition and laser frequencies. 
To simplify the language we will assume this setting for convenience in the following, noting that our reasoning will still
pertain to any other two-level system such as a spin-$\frac{1}{2}$ particle or a Bose-Einstein condensate on an accelerated optical lattice \cite{BEC_lattice}. In other settings, $\Omega(t)$ and $\delta_{2}(t)$ will correspond to different physical quantities.

The goal is to achieve perfect population inversion in a short time in a
two-level quantum system. The system should start at $t=0$ in the ground state and end in the excited state (up to a phase) at final time $T$.
In order to design a scheme to achieve this goal i.e. to design a STA,
we make use of Lewis-Riesenfeld invariants \cite{LR69}. A Lewis-Riesenfeld invariant of $H_{2L}$
is a Hermitian Operator $I\left(t\right)$ such that 
\begin{equation}
\frac{\partial I}{\partial t}+\frac{i}{\hbar}\left[H_{2L},I\right]=0\,.
\end{equation}
In this case $I\left(t\right)$ is given by 
\begin{equation}
I\left(t\right)=\frac{\hbar}{2}\mu\left(\begin{array}{cc}
\cos\left(\theta\left(t\right)\right) & \sin\left(\theta\left(t\right)\right)e^{-i\alpha\left(t\right)}\\
\sin\left(\theta\left(t\right)\right)e^{i\alpha\left(t\right)} & -\cos\left(\theta\left(t\right)\right)
\end{array}\right)
\end{equation}
where $\mu$ is an arbitrary constant with units of frequency to keep $I\left(t\right)$ with dimensions of energy.
The functions $\theta(t)$ and $\alpha(t)$ must satisfy the
following equations:
\begin{eqnarray}
\dot{\theta}&=&\Omega_{I}\cos\alpha-\Omega_{R}\sin\alpha,\label{eq:4}\\
\dot{\alpha}&=&-\delta_{2}-\cot\theta\left(\Omega_{R}\cos\alpha+\Omega_{I}\sin\alpha\right)\,.\label{eq:5}
\end{eqnarray}
The eigenvectors of $I\left(t\right)$ are 
\begin{eqnarray}
\left|\phi_{+}\left(t\right)\right\rangle &=&\left(\begin{array}{c}
\cos\left(\theta/2\right)e^{-i\alpha/2}\\
\sin\left(\theta/2\right)e^{i\alpha/2}
\end{array}\right)\,,\\
\left|\phi_{-}(t)\right\rangle &=&\left(\begin{array}{c}
\sin\left(\theta/2\right)e^{-i\alpha/2}\\
-\cos\left(\theta/2\right)e^{i\alpha/2}
\end{array}\right)
\end{eqnarray}
with eigenvalues $\pm\frac{\hbar}{2}\mu$. One can write a general
solution of the Schr\"{o}dinger equation
\begin{eqnarray}
i\hbar \frac{d}{dt} \left|\Psi\left(t\right)\right\rangle = H_{2L}(t) \left|\Psi\left(t\right)\right\rangle
\end{eqnarray}
as a linear combination of the eigenvectors of $I\left(t\right)$
i.e. $\left|\Psi\left(t\right)\right\rangle =c_{+}e^{i\kappa_{+}(t)}\left|\phi_{+}\left(t\right)\right\rangle +c_{-}e^{i \kappa_{-}(t)}\left|\phi_{-}(t)\right\rangle $
where $c_{\pm}\in\mathbb{C}$ and $\kappa_{\pm}\left(t\right)$ are the Lewis-Riesenfeld
phases \cite{LR69}
\begin{equation}
\dot{\kappa}_{\pm}\left(t\right)=\frac{1}{\hbar}\left\langle \phi_{\pm}\left(t\right)\right.\left|\left(i\hbar\partial_{t}-H_{2L}\left(t\right)\right)\right|\left.\phi_{\pm}\left(t\right)\right\rangle\,.
\label{kappa}
\end{equation}
Therefore, it is possible to construct a solution 
\begin{eqnarray}
\left|\psi\left(t\right)\right\rangle
&=&\left|\phi_{+}\left(t\right)\right\rangle
e^{-i\gamma\left(t\right)/2}
\end{eqnarray}
where $\gamma = \pm 2 \kappa_{\pm}$.
From Eq. \eqref{kappa} we get 
\begin{equation}
\dot{\gamma}=\frac{1}{\sin\theta}\left(\Omega_{R}\cos\alpha+\Omega_{I}\sin\alpha\right)\,.\label{eq:11}
\end{equation}
For population inversion it must be the case that $\theta\left(0\right)=0$
and $\theta\left(T\right)=\pi$. This ensures that $\left|\psi\left(0\right)\right\rangle =\left|1\right\rangle$ and $\left|\psi\left(T\right)\right\rangle =\left|2\right\rangle$ up to a phase. Note, that this method is not limited to going from the ground state to the excited
state; the initial and final states can be determined by changing
the boundary conditions on $\theta$ and $\alpha$. Using Eqs. \eqref{eq:4},
\eqref{eq:5} and \eqref{eq:11} we can retrieve the physical quantities:
\begin{eqnarray}
\Omega_{R}&=&\cos\alpha\sin\theta\,\dot{\gamma}-\sin\alpha\,\dot{\theta}\label{eq:12}\,,\\
\Omega_{I}&=&\sin\alpha\sin\theta\,\dot{\gamma}+\cos\alpha\,\dot{\theta}\label{eq:13}\,,\\
\delta_{2}&=&-\cos\theta\,\dot{\gamma}-\dot{\alpha}\label{eq:14}\,.
\end{eqnarray}
From this we can see that if the functions $\alpha,\gamma$, and $\theta$
are chosen with the appropriate boundary conditions, perfect population inversion would be achieved at a time $T$ assuming no perturbation
or unwanted transitions. These functions will henceforth be referred
to as ancillary functions. In the following section we assume that
there is an additional unwanted coupling to a third level.

%
%


\section{Two-level quantum system with unwanted transition\label{sect3}}

\subsection{Model}

We assume there are in fact three levels in the atom
as shown in Fig. \ref{fig_1_basics}(b) and the energy of level
$\left|j\right\rangle $ is $\hbar\omega_{j}$ where $j=1,2,3$. Without loss of generality
we set $\omega_{1}=0$. The frequency of the laser coupling levels $\left|1\right\rangle $ and $\left|2\right\rangle $
is denoted by $\omega_{L}$. The detuning with the second level is given by 
\begin{equation}
\delta_{2}=\omega_2 - \omega_{L}\, .
\end{equation}
We assume that this laser is also unintentionally coupling levels $\left|1\right\rangle$ and $\left|3\right\rangle$. 
With this in mind, we assume that the Rabi frequency $\Omega_{13}(t)$ differs from $\Omega_{12}(t)$ by a constant complex
number, i.e. 
\begin{equation}
\Omega_{13}\left(t\right)= \beta e^{i \zeta} \Omega_{12} (t)
\end{equation}
where $\zeta,\beta$ are real unknown constants, $\beta \ll 1$. $\Omega_{12}(t)$ is the Rabi frequency coupling levels $\left|1\right\rangle$ and $\left|2\right\rangle$.

A possible motivation for these assumptions
in a quantum-optics setting might be the following:
assume that one needs right circularly polarized light ($\sigma^{+}$) in order to couple states
$\left|1\right\rangle $ and $\left|2\right\rangle $ and one needs left
circularly polarized light ($\sigma^{+}$) to couple states
$\left|1\right\rangle $ and $\left|3\right\rangle $. If the laser light is
-instead of exactly right polarized- elliptically polarized, this
would cause unwanted transitions to level $\left|3\right\rangle$.
Other motivations for these assumptions are possible, especially in other 
quantum systems (different from the quantum-optics setting of an atom
and a classical laser).
Note, that these assumptions are also used in \cite{genov_2013} with the only difference that in that paper a controllable, time-dependent $\zeta$ has been assumed.

The three levels of our atom should have the following state representation:
\begin{eqnarray}
\left|1\right\rangle =\left(\begin{array}{c}
1\\
0\\
0
\end{array}\right)\,,\,
\left|2\right\rangle =\left(\begin{array}{c}
0\\
1\\
0
\end{array}\right)\,,\,
\left|3\right\rangle =\left(\begin{array}{c}
0\\
0\\
1
\end{array}\right)\,.
\end{eqnarray}
Hence our Hamiltonian for the three-level system is 
\begin{align}
H\left(t\right) & =\frac{\hbar}{2}\left(\begin{array}{ccc}
-\delta_{2}(t) & \Omega_{12}^{*}\left(t\right) & \beta e^{-i\zeta}\Omega_{12}^{*}\left(t\right)\\
\Omega_{12}\left(t\right) & \delta_{2}(t) & 0\\
\beta e^{i\zeta}\Omega_{12}\left(t\right) & 0 & -2\Delta+\delta_{2}\left(t\right)
\end{array}\right)\label{eq:original H}
\end{align}
where $\Delta = \omega_2-\omega_3$ is the frequency difference between level $\left|2\right\rangle$ and $\left|3\right\rangle$.
The phase $\zeta$ can be absorbed in a redefinition of the basis state for the
third level and therefore in the following we will just set  $\zeta=0$.

Using the formalism presented in Sect. \ref{sect2}, we can construct schemes which
result in full population inversion in the case of no unwanted transition.
There is a lot of freedom in choosing the ancillary functions. The goal will be
to find the schemes which are very robust against unwanted transitions,
i.e. schemes which result in a nearly perfect population inversion even in the
presence of an unwanted transition.


\subsection{Transition sensitivity\label{sect_q}}

We can write solutions
of the time-dependent Schr\"{o}dinger equation for the Hamiltonian in Eq. \eqref{eq:original H}
if $\beta=0$ as follows
\begin{eqnarray}
\left|\psi_{0}\left(t\right)\right\rangle &=&\left(\begin{array}{c}
\cos\left(\theta/2\right)e^{-i\alpha/2}\\
\sin\left(\theta/2\right)e^{i\alpha/2}\\
0
\end{array}\right)e^{-i\gamma/2}\,,\\
\left|\psi_{1}\left(t\right)\right\rangle &=&\left(\begin{array}{c}
\sin\left(\theta/2\right)e^{-i\alpha/2}\\
-\cos\left(\theta/2\right)e^{i\alpha/2}\\
0
\end{array}\right)e^{i\gamma/2}\,,\\
\left|\psi_{2}\left(t\right)\right\rangle &=&\left(\begin{array}{c}
0\\
0\\
e^{-i\Gamma\left(t\right)}
\end{array}\right)\label{hamiltonian_3l}
\end{eqnarray}
where  $\dot{\Gamma}=\frac{1}{2}\left(-2\Delta+\delta_{2}\right)$.
These solutions form an orthonormal basis at every time $t$.
The ancillary functions $\theta, \alpha, \gamma$ must fulfill Eqs. (\ref{eq:4}),   (\ref{eq:5})
and (\ref{eq:11}).

This unwanted coupling to the third level can be regarded as a perturbation using the approximation that $\beta$
is small. We can write our Hamiltonian \eqref{hamiltonian_3l} as 
\begin{equation}
H\left(t\right)=H_{0}\left(t\right)+\beta V\left(t\right)
\end{equation}
where $\beta$ is the strength of the perturbation,
\begin{equation}
H_{0}\left(t\right)=\frac{\hbar}{2}\left(\begin{array}{ccc}
-\delta_{2}(t) & \Omega_{12}^{*}\left(t\right) & 0\\
\Omega_{12}\left(t\right) & \delta_{2}(t) & 0\\
0 & 0 & -2\Delta+\delta_{2}(t)
\end{array}\right)
\end{equation}
 and
\begin{equation}
V\left(t\right)=\frac{\hbar}{2}\left(\begin{array}{ccc}
0 & 0 & \Omega_{12}^{*}\left(t\right)\\
0 & 0 & 0\\
\Omega_{12}\left(t\right) & 0 & 0
\end{array}\right)\,.\label{pot}
\end{equation}
Using time-dependent perturbation theory we can calculate the  probability of being in state $\left|2\right\rangle$ at time $T$ as
\begin{equation}
P_{2}=1- \beta^{2} q + \mathcal{O}\left(\beta^{4}\right)\,
\end{equation}
where
\begin{eqnarray}
q = \frac{1}{\hbar^2} \sum_{k=0}^2 \fabsq{\int_0^T dt\, \la
  \psi_0(t)|V(t)|\psi_k(t) \ra}\,.
\end{eqnarray}
If we substitute in the expression for the perturbation \eqref{pot} then we get
\begin{eqnarray}
q &=& \frac{1}{4} \left|\int_{0}^{T}dt\,  \cos\left(\frac{\theta}{2}\right)
\left(\sin\theta \, \dot\gamma - i \dot\theta\right) e^{i F(t)+i\Delta t} \right|^2\nonumber\\
&=& \left|\int_{0}^{T}dt\, \frac{d}{dt} \left[
\sin\left(\frac{\theta(t)}{2}\right) e^{i F(t)} \right] e^{i\Delta t} \right|^2
\label{eq_q}
\end{eqnarray}
where $F(t)=\frac{1}{2} \int_0^t ds\, (1+\cos\theta(s)) \dot\gamma(s)$.
The $q$ quantifies how sensitive a given protocol
(determined by the ancillary functions) is concerning the unwanted
transition to level $\left|3\right\rangle$ . Therefore we will call $q$ {\it transition
  sensitivity} in the following.
Our goal will be to determine protocols or schemes which would maximize
$P_{2}$ or equivalently minimize $q$.


\subsection{General properties of the transition sensitivity\label{sect_2level_properties}}

We will begin by examining some general properties of the transition sensitivity
$q$. First, we note that $q$ is always independent of $\alpha$. In the case where $\dot{\gamma}=0$ the transition sensitivity is symmetric
about $\Delta \leftrightarrow -\Delta$.

In the case of $\Delta = 0$, the integral in Eq. \eqref{eq_q}
can be easily evaluated by taking into account that $\theta(T)=\pi$ and
$\theta(0)=0$. From this we see that
\begin{eqnarray}
q=1 \; \mbox{if} \; \Delta = 0\,.
\end{eqnarray}
This means there is no possibility in the case of $\Delta=0$ to completely
reduce the influence of the unwanted transition.

In the following, we will show that even for $|\Delta| < 1/T$ the transition
probability $q$ cannot be zero.
By partial integration, we get
\begin{eqnarray}
q = \left|1 - i \Delta M\right|^2= 1 + 2\Delta \mbox{Im}(M) + \Delta^2 \fabsq{M}
\end{eqnarray}
where
\begin{eqnarray}
M &=&\int_0^T dt\, \sin\left(\frac{\theta(t)}{2}\right)
\times \nonumber\\
& & \fexp{i (t-T)\Delta -\frac{i}{2} \int_t^T ds\,
  (1+\cos\theta(s)) \dot\gamma(s)}\,.\nonumber\\
\end{eqnarray}
We have $q \ge (1 + \Delta \mbox{Im}(M))^2$ and
\begin{eqnarray}
|\mbox{Im}(M)|
&\le&  \int_0^T dt\,
\left|\sin\left(\frac{\theta(t)}{2}\right)\right| \le T\,.
\end{eqnarray}
Let us assume $|\Delta|T < 1$ then
\begin{eqnarray}
q &\ge& (1 - |\Delta| |\mbox{Im}(M)|)^2\nonumber\\
&\ge& \left(1 - |\Delta| \int_0^T dt\,
\left|\sin\left(\frac{\theta(t)}{2}\right)\right|\right)^2\nonumber\\
&\ge& (1- |\Delta| T)^2\,.
\label{bound_q}
\end{eqnarray}
So we get $q > 0$ if $|\Delta|T < 1$, i.e.
this means that a necessary condition for $q = 0$ is $T \ge 1/|\Delta|$.

The next question which we will address is whether there could be a scheme
(independent of $\Delta$) which results in $q=0$ for all $|\Delta| > 1/T$.
For this we would need
\begin{eqnarray}
H(\Delta) := \int_{0}^{T}dt\, \frac{d}{dt} \left[
G\left(t\right)\right] e^{i\Delta t} \stackrel{!}{=} 0
\label{cond}
\end{eqnarray}
for all $|\Delta| > 1/T$, where $G(t)= \fsin{\theta(t)/2} e^{i F(t)}$.
The left-hand side of this equation, $H(\Delta)$, is simply the Fourier transform of
$h(t)=\chi_{[0,T]}(t)\frac{d}{dt} \left[G\left(t\right) \right]$
(where $\chi_{[0,T]}(t) = 1$ for $0 \le t \le T$ and zero
otherwise). $h(t)$ has compact support. If Eq. \eqref{cond} would be true then this would
mean that the Fourier transform $H(\Delta)$
of the compactly supported function $h(t)$ also has compact support. 
This is not possible and therefore there can be no
($\Delta$-independent) protocol which results in $q=0$ for all $|\Delta| >
1/T$. Nevertheless, we will show below that for a fixed $\Delta$ there are schemes
resulting in $q=0$.

It is also important to examine general properties for $|\Delta| \gg
1/T$. From the previous remark (and the property that a Fourier transform of
any function vanishes at infinity) it is immediately clear that we get
$q \to 0$ for $|\Delta| \to \infty$. Using partial integration we can derive a series expansion of $q$ in
$1/\Delta$. We use
\begin{eqnarray}
\int_0^T dt \dot G (t) e^{i \Delta t} &=& -\frac{i}{\Delta} \left[\dot G (t)
  e^{i \Delta t}\right]_0^T + \frac{i}{\Delta} \int_0^T dt \ddot G(t) e^{i
  \Delta t}\nonumber\\
 &=& -\frac{i}{\Delta} \left[\dot G (t)
  e^{i \Delta t}\right]_0^T + o\left(\frac{1}{\Delta}\right).
\end{eqnarray}
Hence, in the case where $|\Delta| \gg 1/T$ the transition sensitivity is
\begin{eqnarray}
q = \frac{1}{\Delta^2} \frac{1}{4} \dot \theta (0)^2 + ...
\end{eqnarray}
where we have taken into account that $\theta(0)=0$ and $\theta(T)=\pi$.
By repeating partial integration, we get the higher orders in this
$1/\Delta$ series.

If we demand
\begin{eqnarray}
\dot\theta(0)=\dot\theta(T)=\ddot\theta(0)=0
\label{additional}
\end{eqnarray}
then this first term and the next terms in the $1/\Delta$ series expansion
of the transition sensitivity vanish. The first non-vanishing term is now
\begin{eqnarray}
q=\frac{1}{\Delta^6} \dddot \theta(0)^2 + ...
\end{eqnarray}

\subsection{Reference case: flat $\pi$ pulse}

As a reference case we will consider a flat $\pi$ pulse with
\begin{eqnarray}
\Omega_R = -\frac{\pi}{T} \sin \alpha,\: \Omega_I = \frac{\pi}{T} \cos\alpha
\end{eqnarray}
with a constant phase $\alpha$.
This scheme corresponds to $\theta(t) = \pi \frac{t}{T}$ and $\gamma(t)=0$.

\begin{figure}[t]
\begin{center}
(a) \includegraphics[angle=0,width=0.8\linewidth]{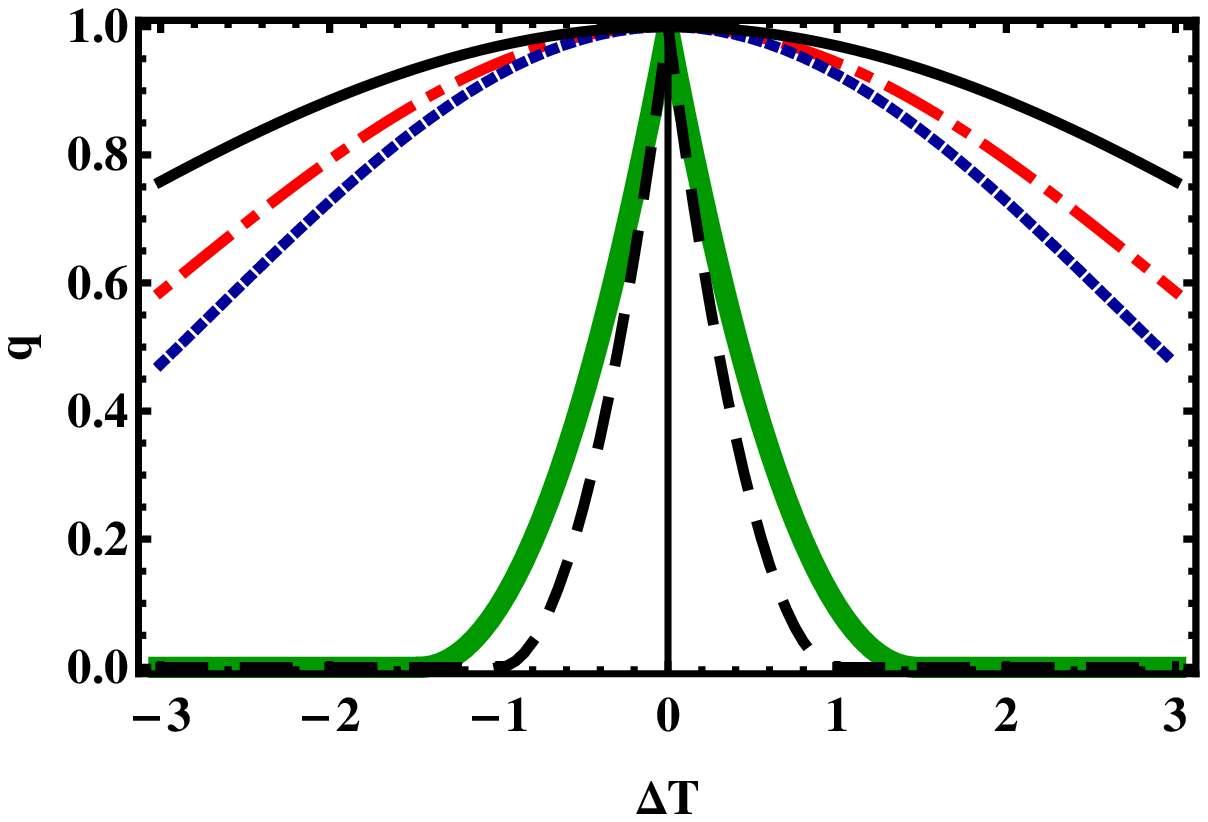}

(b) \includegraphics[angle=0,width=0.8\linewidth]{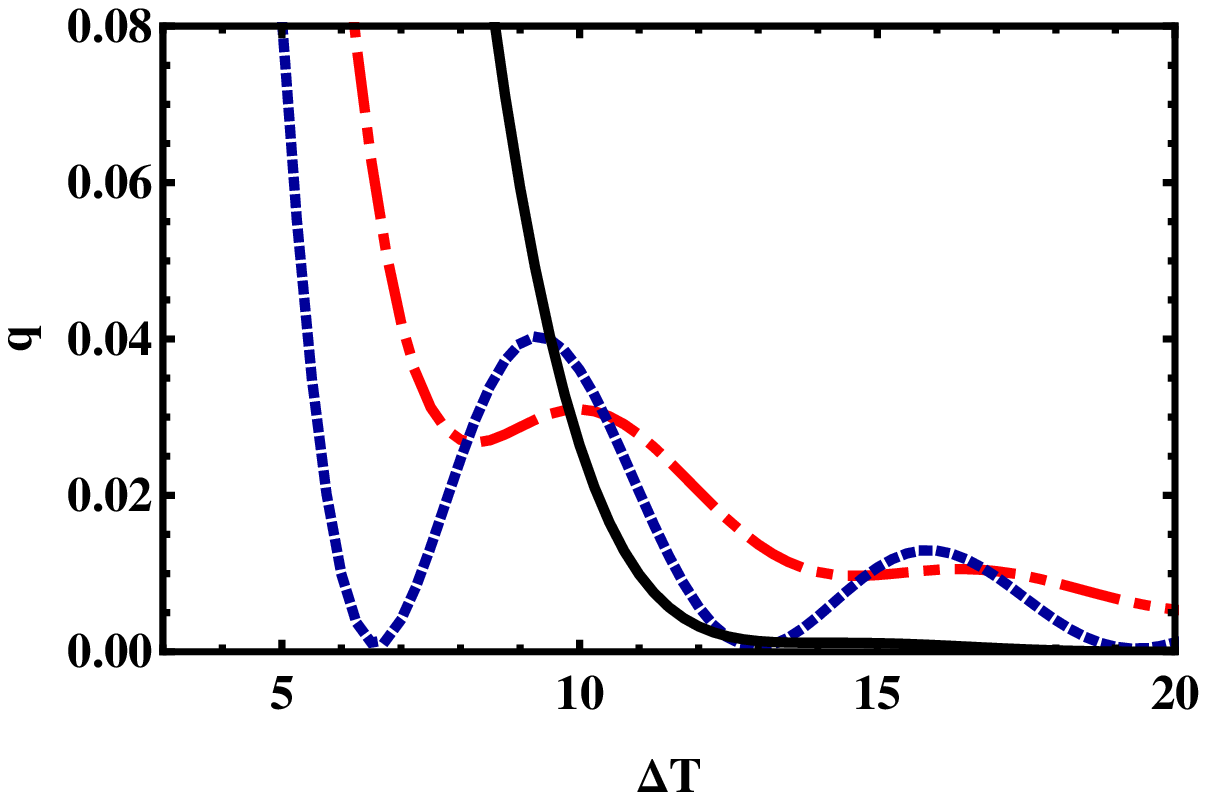}
\end{center}
\caption{\label{fig_2_q} (Color online) Transition sensitivity $q$ versus
   $\Delta T$ for different schemes; reference case of a flat $\pi$ pulse
  (red, dot-dashed line); other $\pi$ pulse given by Eq. \eqref{otherpi} (blue, dotted
  line); scheme given by Eq. \eqref{large_delta_scheme} (black, solid line);
also in (a): scheme in Eq. \eqref{scheme_num_3l} with numerically optimized parameters $c_0$ and
$c_1$ (green, thick, solid line); lower bound for $q$ as in Eq. \eqref{bound_q}
(black, dashed line).}  
\end{figure}

The transition sensitivity can be easily calculated
\begin{eqnarray}
q &=& \left|\int_{0}^{T}dt\, \frac{d}{dt} \left[
\sin\left(\frac{\pi t}{2 T}\right) \right] e^{i\Delta t} \right|^2\nonumber\\
&=& \frac{\pi ^2 \left(4 \Delta^2 T^2-4 \pi  \Delta T \sin
   (\Delta T)+\pi ^2\right)}{\left(\pi ^2-4 \Delta^2
   T^2\right)^2}\,.
\label{eq_q_pi}
\end{eqnarray}
This transition sensitivity $q$ is plotted in Fig. \ref{fig_2_q}(a) and (b).
It can be seen that $q$ is one for $\Delta = 0$ and it goes to zero for large
$|\Delta|$  as is expected. The transition sensitivity for
the flat $\pi$ pulse is never exactly zero.

%
%

%
\begin{figure}[t]
\begin{center}
\includegraphics[angle=0,width=0.8\linewidth]{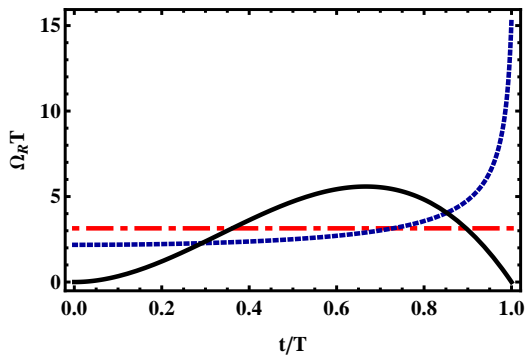}
\end{center}
\caption{\label{fig_3_rabi} (Color online) Rabi frequencies $\Omega_R(t)$ versus time for
  different scheme:
reference case of a flat $\pi$ pulse (red, dot-dashed, line);
other $\pi$ pulse given by Eq. \eqref{otherpi} (blue, dotted line);
scheme given by Eq. \eqref{large_delta_scheme} (black, solid line).}  
\end{figure}
%

\subsection{Other examples of $\pi$ pulses}

Let us examine two other examples of protocols.
Suppose $\gamma\left(t\right)=0$,
$\theta\left(t\right)=2\arcsin\left(\frac{t}{T}\right)$.
Then we get
\begin{equation}
q=\left|\frac{\left(1-e^{i\Delta T}\right)}{\Delta T}\right|^{2}\, .
\end{equation}
In order to achieve $q=0$ one must have
$T=\frac{2n\pi}{\Delta}$. We also set $\alpha$ constant and then
the associated physical quantities for this protocol are 
\begin{equation}
\delta_{2}\left(t\right)=0\,,\, \:\Omega_{12}\left(t\right)=\frac{2ie^{i\alpha}}{T\sqrt{1-\frac{t^{2}}{T^{2}}}}\,.
\end{equation}
This is a type of $\pi$ pulse. Unfortunately the Rabi frequency $\Omega_{12}$
diverges at $t=T$. To stop divergence we set
\begin{eqnarray} 
\theta\left(t\right)=\frac{\pi}{\arcsin\left(1-\epsilon\right)}\arcsin\left(\left(1-\epsilon\right)\frac{t}{T}\right)\,
\end{eqnarray}
where $0<\epsilon \ll 1$. By setting $\alpha=-\pi/2$ the corresponding Rabi frequency is real (i.e. $\Omega_I(t)=0$) and 
\begin{eqnarray}
\Omega_R (t)=
\frac{\pi\left(1-\epsilon\right)}{\arcsin\left(1-\epsilon\right)T\sqrt{1-\frac{t^{2}\left(\epsilon-1\right)^{2}}{T^{2}}}}\,.
\label{otherpi}
\end{eqnarray}
It also follows that $\delta_{2}=0$.
The corresponding transition sensitivity with $\epsilon=0.01$ is also plotted
in Fig. \ref{fig_2_q}(a) and (b). Note that this scheme converges for $\epsilon \to 1$ to a flat $\pi$ pulse.

We also construct a scheme fulfilling Eqs. \eqref{additional} which results in a low $q$
value for large $|\Delta|$. For this scheme we set 
\begin{eqnarray}
\theta(t) =-\frac{3 \pi  t^4}{T^4} + \frac{4 \pi  t^3}{T^3}
\label{large_delta_scheme}
\end{eqnarray}
and $\gamma=0$. The corresponding transition probability can be seen
in Fig. \ref{fig_2_q}(a) and (b). The transition sensitivity for this scheme is lower than that of the flat $\pi$-pulse for $\Delta T > 10$, meaning it is less sensitive to unwanted transitions.
If we set $\alpha=-\pi/2$ then we get $\Omega_R(t)=\frac{12\pi t^{2}(T-t)}{T^{4}}$,
$\Omega_I=0$ and $\delta_2 = 0$.

\subsection{Numerically optimized scheme with $q = 0$}

In the following we will present an example of a class of schemes which
can be optimized to achieve a zero transition sensitivity for a fixed $\Delta$.
We use the ansatz
\begin{eqnarray}
\begin{array}{rcl}
\gamma(t) &=& c_0 \theta(t)\,,\\
\theta(t) &=& (\pi-c_1) t/T + c_1 t^3/T^3
\end{array}
\label{scheme_num_3l}
\end{eqnarray}
where the parameters $c_0$ and $c_1$ were numerically calculated in order to 
minimize $q$ for a given $\Delta$.
The result is shown in Fig. \ref{fig_2_q}(a).
As it can be seen, 
we can construct schemes which make $q$ vanish
for $|\Delta| T \ge 1.5$.

\begin{figure}[t]
\begin{center}
(a) \includegraphics[angle=0,width=0.8\linewidth]{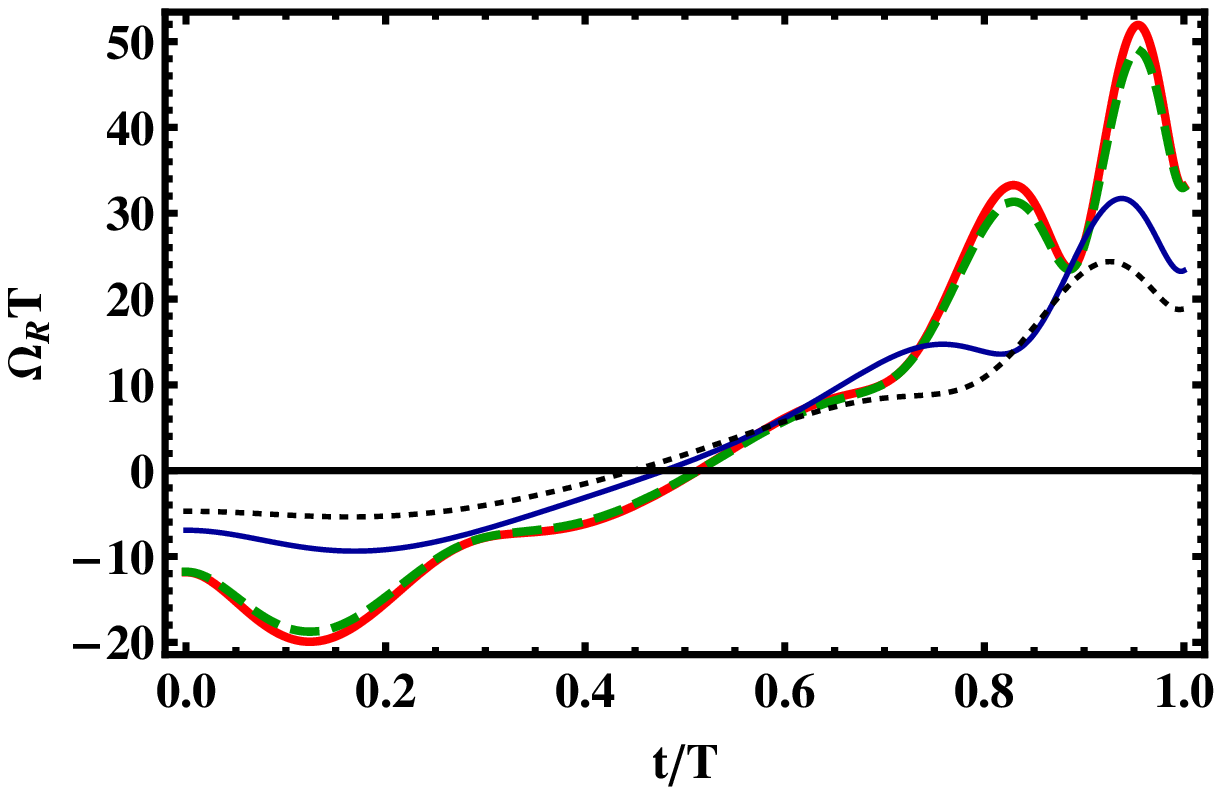}

(b) \includegraphics[angle=0,width=0.8\linewidth]{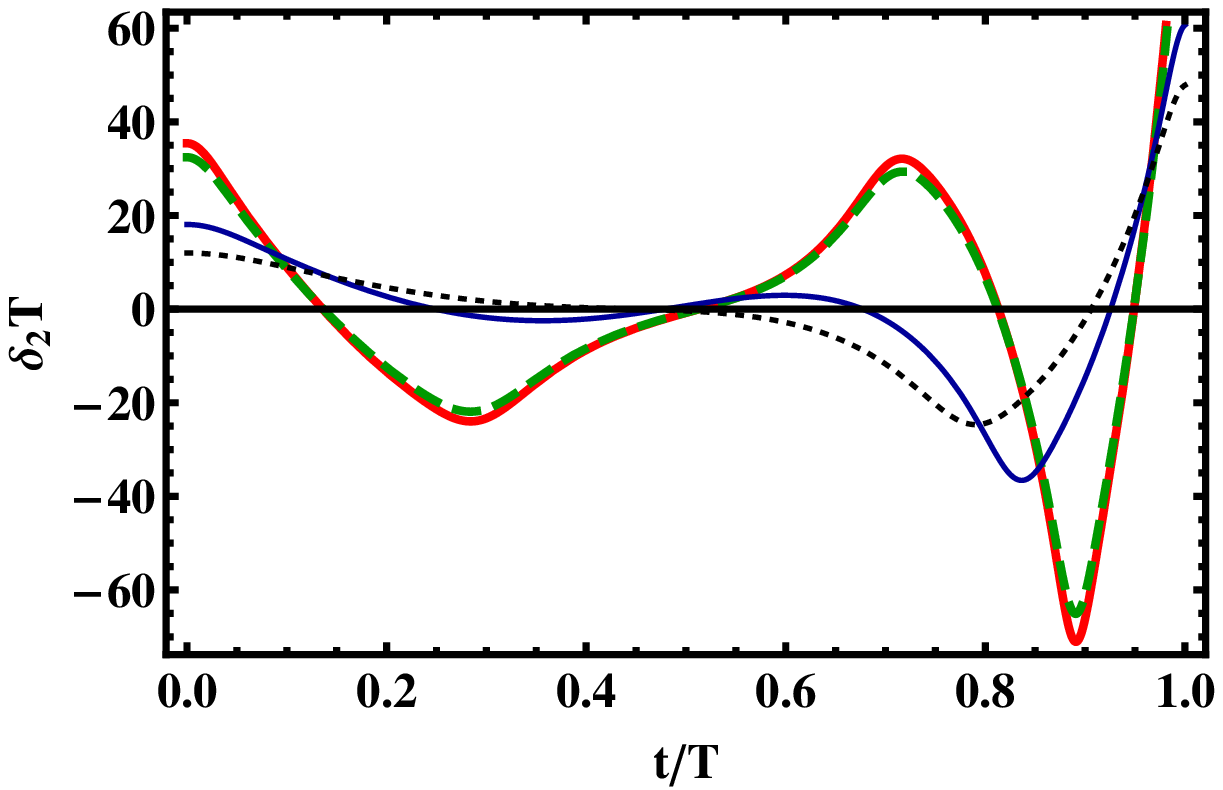}
\end{center}
\caption{\label{fig_4} (Color online) Physical potentials for the numerically optimized
  schemes in Eq. \eqref{scheme_num_3l} versus time $t$.
(a) Rabi frequency $\Omega_R$; (b) detuning $\delta_{2}$.
$\Delta T = 0.2$ (red, thick, solid line),
$\Delta T = 1.0$ (green, dashed line),
$\Delta T = 2.0$ (blue, thin, solid line),
$\Delta T = 3.0$ (black, dotted line).}  
\end{figure}

$\alpha(t)$ is chosen so that the Rabi frequency is real. The corresponding Rabi frequency $\Omega_R$ and the
detuning $\delta_2$ is shown in Fig. \ref{fig_4}
for different values of $\Delta T$.

Note that we pick the ansatz \eqref{scheme_num_3l} because it is
simple. It is still possible to optimize the ansatz further for
example with the goal of minimizing the maximal Rabi frequency. Moreover, the
ansatz could be modified so that the Rabi frequency is
zero at initial and final times.

%
%

\subsection{Comparison of the transition probability}

\begin{figure}[t]
\begin{center}
(a) \includegraphics[angle=0,width=0.7\linewidth]{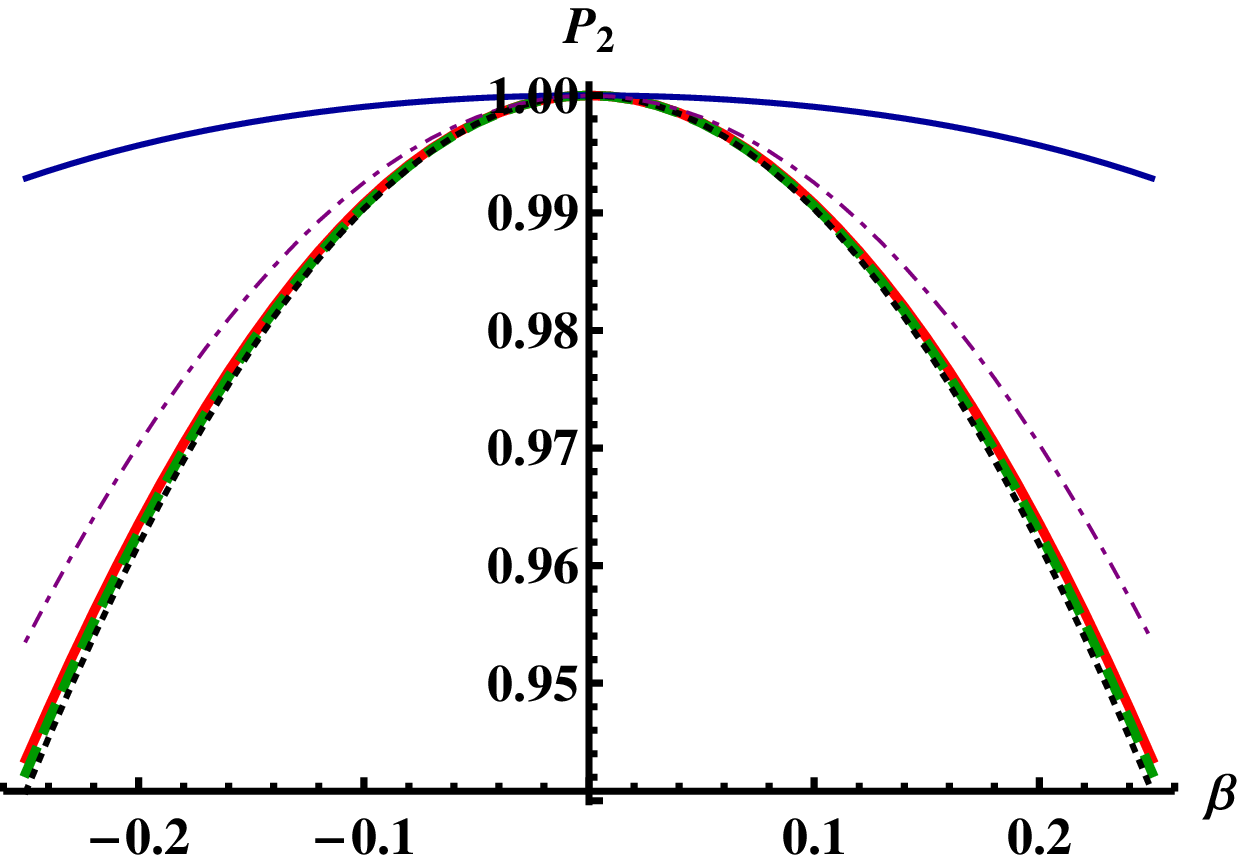}

(b) \includegraphics[angle=0,width=0.7\linewidth]{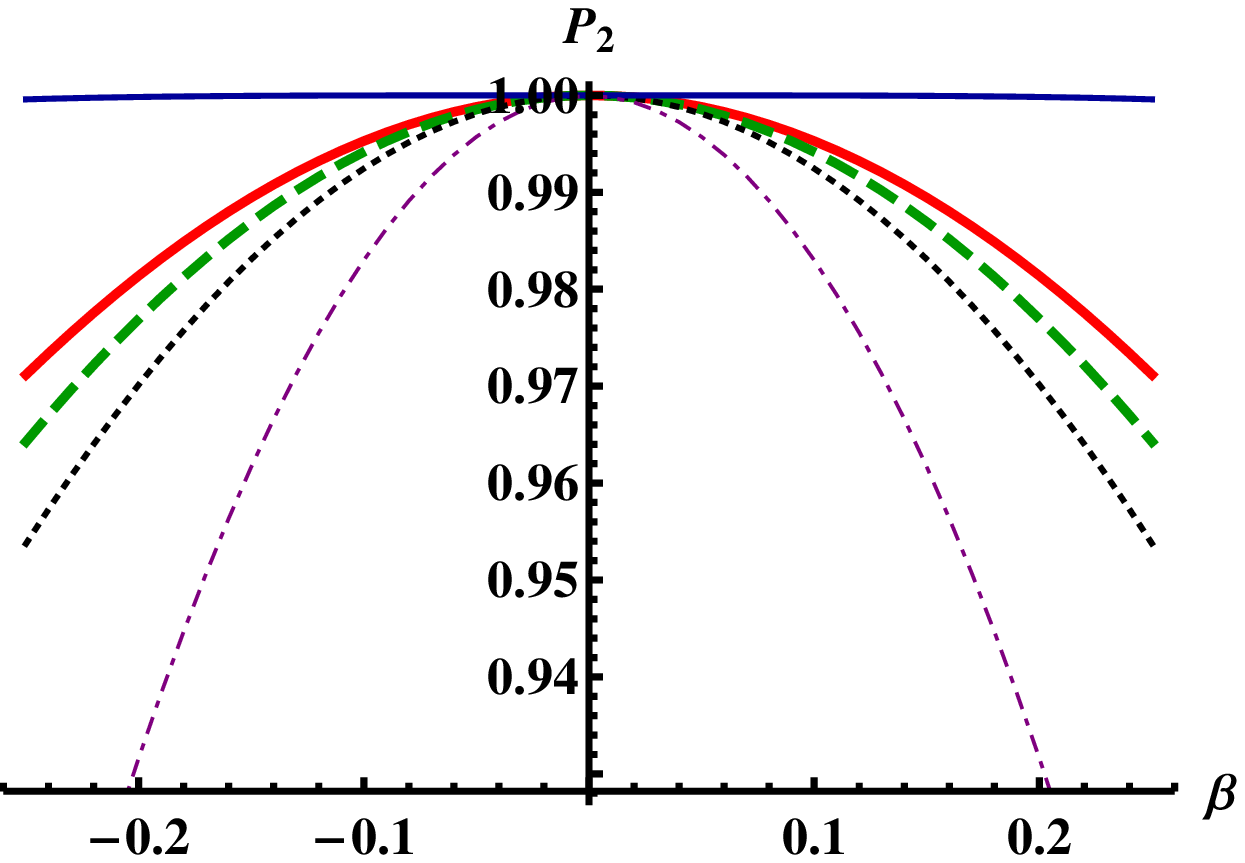}
\end{center}
\caption{\label{fig_6_p2} (Color online) Transition probability $P_2$ versus perturbation
  strength $\beta$ for different schemes:
reference case of a flat $\pi$ pulse (green, thick, dashed line);
other $\pi$ pulse given by Eq. \eqref{otherpi}
(red, thick, solid line);
scheme given by Eq. \eqref{large_delta_scheme} (black, thin, dashed line);
scheme in Eq. \eqref{scheme_num_3l} with numerically
optimized parameters $c_0$ and $c_1$ (blue, thin, solid line);
adiabatic scheme (purple, thin, dot-dashed line);
(a) $\Delta T = 1.0$, (b) $\Delta T = 3.0$.}  
\end{figure}

In the following we compare the effectiveness of the different schemes. To do this we
compare the exact (numerically calculated) transition probability $P_2$ for
the different schemes versus $\beta$ for different values of $\Delta$.
This can be seen in Fig. \ref{fig_6_p2}. From this we see that the transition sensitivity is a good indicator of a stable scheme.
This is however not the only useful quantity to know about a particular scheme. We also consider the area of the pulse $A :=  \int_0^T dt\,
\sqrt{\Omega_R^2 + \Omega_I^2}$ and its energy $E:=\hbar \int_0^T dt\, \left(\Omega_R^2 +
\Omega_I^2\right)$. The values for the different schemes are shown in
Table \ref{tab1}. It can be seen that the numerically optimized schemes require
a higher energy than three different variations of a $\pi$ pulse.

For completeness we also include the following sinusoidal adiabatic scheme \cite{Lu,Xiao} in our comparison: 
\begin{eqnarray}
\begin{array}{rcl}
\Omega_{12}\left(t\right)&=&\Omega_{0}\sin\left(\frac{\pi t}{T}\right)\,,\\[0.2cm]
\delta_{2}\left(t\right)&=&-\delta_{0}\cos\left(\frac{\pi t}{T}\right)\,.
\end{array}
\end{eqnarray}
We have chosen $\Omega_0$ so that the adiabatic scheme requires the same
energy as the numerically optimized scheme. In addition, we have also
optimized the $\delta_0$ to maximize the value of $P_2$ for the error-free
case $\beta=0$. The energy is high enough that  the adiabatic
scheme results in a nearly perfect population inversion in the error-free
case. Nevertheless, the numerically optimized scheme is less sensitive to
unwanted transitions, i.e. the numerically optimized scheme results in a higher
$P_2$ for non-zero $\beta$.


\begin{table}
\begin{tabular}{|c|c|c|}
\hline 
 & $A [\pi]$ & $E [\pi^{2}\hbar/T]$\\
\hline 
Flat $\pi$ pulse & $1$  & $1$\\
\hline 
Critical timing scheme$\left(\epsilon=0.01\right)$,&&\\
Eq. \eqref{otherpi} & $1$  & $1.28$\\
\hline 
Large $\Delta$ scheme, Eq. \eqref{large_delta_scheme}
& $1$  & $\frac{48}{35}$\\
\hline 
Numerically optimized scheme,& &\\
 Eq. \eqref{scheme_num_3l}& &\\
$\Delta T=1.0\:(c_{0}=1.376, c_{1}=14.927)$ & $4.79$ & $36.56$\\
$\Delta T=3.0\:(c_{0}=1.266, c_{1}=7.873)$ & $2.49$ & $10.51$\\
\hline 
Adiabatic Scheme & $2T\Omega_{0}\pi^{-2}$ & $\frac{1}{2}\pi^{-2}T^{2}\Omega_{0}^{2}$\\
$\Delta T=1.0$ & $5.44$ & $36.56$\tabularnewline
$\Delta T=3.0$ & $2.92$ & $10.51$\tabularnewline
\hline 
\end{tabular}\caption{\label{tab1}Pulse area $A$ and energy $E$ for different protocols.}
\end{table}

%

%
%


\section{Invariant-based shortcuts in three-level systems \label{sect4}}
In this section, we will review the derivation of invariant-based STA in three-level
systems \cite{sta_3level} (for an application see for
example \cite{Tseng_2012}).
We use a different notation than \cite{sta_3level} to underline the connection between the two and three-level Hamiltonians
in Eq. \eqref{eq:1} and Eq. \eqref{3level_H0} respectively (see for example \cite{three-two}). In addition, we will introduce different boundary
conditions for the ancillary functions than those used in \cite{sta_3level}.

We assume our three-level system has a Hamiltonian of the form
\begin{equation}
H_{3L}\left(t\right)=\frac{\hbar}{2}\left(\begin{array}{ccc}
0 & \Omega_{12}\left(t\right) & 0\\
\Omega_{12}\left(t\right) & 0 & \Omega_{23}\left(t\right)\\
0 & \Omega_{23}\left(t\right) & 0
\end{array}\right)
\label{3level_H0}
\end{equation}
where $\Omega_{12}$ and $\Omega_{23}$ are real. This could for example describe a three-level atom with two on resonance lasers (one coupling states $\left|1\right\rangle $ and $\left|2\right\rangle $ and the other coupling states $\left|2\right\rangle $ and $\left|3\right\rangle $). The Lewis-Riesenfeld invariant for this Hamiltonian is
\begin{equation}
I\left(t\right)=\frac{\hbar}{2}\mu\left(\begin{array}{ccc}
0 & -\sin\theta\sin\alpha & -i\cos\theta\\
-\sin\theta\sin\alpha & 0 & -\sin\theta\cos\alpha\\
i\cos\theta & -\sin\theta\cos\alpha & 0
\end{array}\right)
\end{equation}
where $\mu$ is a constant in units of frequency to keep $I\left(t\right)$
in units of energy. The ancillary functions $\alpha\left(t\right)$ and $\theta\left(t\right)$
satisfy 
\begin{eqnarray}
\dot{\theta}&=&\frac{1}{2}\left(\Omega_{12}\cos\alpha-\Omega_{23}\sin\alpha\right)\,,
\label{3level_diff1}\\
\dot{\alpha}&=&-\frac{1}{2}\cot\theta\left(\Omega_{23}\cos\alpha+\Omega_{12}\sin\alpha\right)\,.
\label{3level_diff2}
\end{eqnarray}
Note the similarity with Eqs. \eqref{eq:4} and \eqref{eq:5}. This is due to the aforementioned connection between the two- and three-level Hamiltonians.
The eigenstates of $I\left(t\right)$ are 
\begin{equation}
\left|\phi_{0}\left(t\right)\right\rangle =\left(\begin{array}{c}
-\sin\theta\cos\alpha\\
-i\,\cos\theta\\
\sin\theta\sin\alpha
\end{array}\right)
\end{equation}
and 
\begin{equation}
\left|\phi_{\pm}\left(t\right)\right\rangle =\frac{1}{\sqrt{2}}\left(\begin{array}{c}
\cos\theta\cos\alpha\pm i\,\sin\alpha\\
-i\,\sin\theta\\
-\cos\theta\sin\alpha\pm i\,\cos\alpha
\end{array}\right)
\end{equation}
with eigenvalues $\lambda_{0}=0$ and $\lambda_{\pm}=\pm1$ i.e. $I\left(t\right)\left|\phi_{n}\left(t\right)\right\rangle =\lambda_{n}\left|\phi_{n}\left(t\right)\right\rangle $
and the label $n=0,\pm$. The Lewis-Riesenfeld phases $\kappa_{n}\left(t\right)$ are $\kappa_{0}=0$ and
\begin{equation}
\kappa_{\pm}=\mp\int_{0}^{t}dt^{'}\left(\dot{\alpha}\cos\theta-\frac{1}{2}\left(\Omega_{12}\sin\alpha+\Omega_{23}\cos\alpha\right)\sin\theta\right).
\end{equation}
A solution of the time-dependent Schr\"odinger equation with the Hamiltonian \eqref{3level_H0} is now $|\Psi(t)\ra = |\phi_0 (t) \ra$.
In order for the solution $|\Psi(t)\ra$ to evolve from the
initial state $\left|1\right\rangle $ to the final state $\left|3\right\rangle $
we must impose the following boundary conditions on $\alpha$ and $\theta$:
\begin{eqnarray}
\theta(0)=-\frac{\pi}{2}\,,\,\theta (T)= \frac{\pi}{2}\,,\,
\alpha(0)=0\,,\,\alpha(T)=\frac{\pi}{2}\,.
\label{3l_cond1}
\end{eqnarray}
One could impose the following additional boundary conditions in order
to make the Rabi frequencies have a finite limit at the initial and final times
\begin{eqnarray}
\dot{\alpha}(0)=0\,,\,\dot{\alpha} (T)=0\,,\,
\dot{\theta}(0)\neq0\,,\,\dot{\theta} (T)\neq0\,.
\label{3l_cond2}
\end{eqnarray}
Note that the boundary conditions given by Eqs. \eqref{3l_cond1} and \eqref{3l_cond2} are an alternative choice to the ones imposed in \cite{sta_3level}.

Using Eqs. \eqref{3level_diff1} and \eqref{3level_diff2} we can calculate the Rabi frequencies
\begin{eqnarray}
\Omega_{12}\left(t\right)&=& 2\left(-\dot{\alpha}\tan\theta\sin\alpha+\dot{\theta}\cos\alpha\right)\,,\\
\Omega_{23}\left(t\right)&=& -2\left(\dot{\alpha}\tan\theta\cos\alpha+\dot{\theta}\sin\alpha\right)\, .
\end{eqnarray}
If the functions $\alpha$ and $\theta$ fulfill Eqs. \eqref{3l_cond1} and \eqref{3l_cond2}, then the corresponding Rabi frequencies will lead to full population inversion $\left|1\right\rangle\to \left|3\right\rangle$.


\section{Unwanted transitions in three-level systems \label{sect5}}

\subsection{Model}

Now we assume that there is an unwanted coupling to a fourth level as shown
in Fig. \ref{fig_1_basics}(c). Analogous to Section \ref{sect3}, we assume
that the laser coupling levels $\left|2\right\rangle$ and $\left|3\right\rangle$ also unintentionally couples levels $\left|2\right\rangle$ and $\left|4\right\rangle$ as well.
Hence we assume for the Rabi frequency
\begin{equation}
\Omega_{24}\left(t\right)=\beta e^{i\nu}\Omega_{23}\left(t\right)
\end{equation}
where $\beta,\nu\in\mathbb{R}$ are unknown constants and $\beta\ll1$.
The Hamiltonian for this four-level system is given by
\begin{equation}
H\left(t\right)=\frac{\hbar}{2}\left(\begin{array}{cccc}
0 & \Omega_{12} & 0 & 0\\
\Omega_{12} & 0 & \Omega_{23} & \beta e^{-i\nu}\Omega_{23}\\
0 & \Omega_{23} & 0 & 0\\
0 & \beta e^{i\nu}\Omega_{23} & 0 & -2\Delta
\end{array}\right)
\end{equation}
where $\Delta=\omega_{3}-\omega_{4}$ and $\hbar\omega_{j}$ is the
energy of state $\left|j\right\rangle $.
As in the previous case, one can redefine the state $\left|4\right\rangle $ to remove
the phase. Hence we set $\nu=0$ in the following.

Using the formalism presented in Sect. \ref{sect4}, we can construct schemes which
result in full population inversion in the case of no unwanted transitions.
Again, there is a lot of freedom in choosing the ancillary functions and
the goal will be to find the schemes which are stable concerning these
unwanted transitions.

\subsection{Transition sensitivity}

We once again regard this unwanted transition as a perturbation. To treat it as such we write the Hamiltonian as 
\begin{equation}
H\left(t\right)=H_{0}\left(t\right)+\beta V\left(t\right)
\end{equation}
where 
\begin{equation}
H_{0}\left(t\right)=\frac{\hbar}{2}\left(\begin{array}{cccc}
0 & \Omega_{12} & 0 & 0\\
\Omega_{12} & 0 & \Omega_{23} & 0\\
0 & \Omega_{23} & 0 & 0\\
0 & 0 & 0 & -2\Delta
\end{array}\right)
\end{equation}
and
\begin{equation}
V\left(t\right)=\frac{\hbar}{2}\left(\begin{array}{cccc}
0 & 0 & 0 & 0\\
0 & 0 & 0 & \Omega_{23}\\
0 & 0 & 0 & 0\\
0 & \Omega_{23} & 0 & 0
\end{array}\right)\,.
\end{equation}
If $\beta=0$
then the time-dependent Schr\"{o}dinger equation for $H\left(t\right)$ has the following set of orthonormal solutions:
\begin{eqnarray}
\left|\psi_{0}\left(t\right)\right\rangle & =&\left(\begin{array}{c}
-\sin\theta \cos\alpha\\
-i\,\cos\theta\\
\sin\theta\sin\alpha\\
0
\end{array}\right)e^{i\,\kappa_{0}}\,,
\end{eqnarray}
\begin{eqnarray}
\left|\psi_{1}\left(t\right)\right\rangle &=&\frac{1}{\sqrt{2}}\left(\begin{array}{c}
\cos\theta\cos\alpha+i\,\sin\alpha\\
-i\,\sin\theta\\
-\cos\theta\sin\alpha+i\,\cos\alpha\\
0
\end{array}\right)e^{i\,\kappa_{+}}\,,
\end{eqnarray}
\begin{eqnarray}
\left|\psi_{2}\left(t\right)\right\rangle &=&\frac{1}{\sqrt{2}}\left(\begin{array}{c}
\cos\theta\cos\alpha-i\,\sin\alpha\\
-i\,\sin\theta\\
-\cos\theta\sin\alpha-i\,\cos\alpha\\
0
\end{array}\right)e^{i\,\kappa_{-}}\,,
\end{eqnarray}
\begin{eqnarray}
\left|\psi_{3}\left(t\right)\right\rangle &=&\left(\begin{array}{c}
0\\
0\\
0\\
e^{i\Delta t}
\end{array}\right)\,.
\end{eqnarray}

Using time-dependent perturbation theory similar to Section \ref{sect_q}, we get
for the probability $P_3$ to end in the state $\left|3\right\rangle$ at time $t=T$ that
\begin{equation}
P_{3}=1-\beta^{2}Q +\mathcal{O}\left(\beta^{4}\right)\,.
\end{equation}
where
\begin{eqnarray}
Q &=&\left|\int_0^Tdt\, e^{i\Delta t}
\left(\dot{\alpha}\sin\theta\cos\alpha + \dot{\theta}\cos\theta\sin\alpha\right)\right|^{2}\nonumber\\
&=&\left|\int_0^Tdt\, e^{i\Delta t}
\frac{d}{dt}\left(\sin\theta\sin\alpha\right)\right|^{2}\,.
\label{def_Q}
\end{eqnarray}
Similar to Sect. \ref{sect3}, the $Q$ quantifies how sensitive a given protocol is concerning the unwanted
transition to level $\left|4\right\rangle$. As before we will call $Q$ {\it transition
  sensitivity} in the following and our goal will be to determine protocols or schemes which would
minimize $Q$.

\subsection{General properties of the transition sensitivity}

We start by examining some general properties of the transition
sensitivity $Q$ given by Eq. \eqref{def_Q} by noting that $Q$ is independent of the
sign of $\Delta$. By taking into account the boundary
conditions for $\theta(t)$ and $\alpha(t)$ we find that 
\begin{eqnarray}
Q=1 \; \mbox{if} \; \Delta = 0\,.
\end{eqnarray}
 
Similar to Sect. \ref{sect_2level_properties}, we get by partial integration
\begin{equation}
Q=\left|1-i\Delta N\right|^{2}=1+2\Delta \mbox{Im}N+\Delta^{2}\left|N\right|^{2}
\end{equation}
where 
\begin{equation}
N=-\int_{0}^{T}dt\, e^{i\Delta\left(t-T\right)}\sin\theta\sin\alpha\,.
\end{equation}
Therefore $Q\geq\left(1+\Delta \mbox{Im}\left(N\right)\right)^{2}$ and 
\begin{align}
\left|\mbox{Im}\left(N\right)\right| & \leq\left|\int_{0}^{T}dt\,\sin\left(\Delta\left(t-T\right)\right)\sin\theta\sin\alpha\right|\nonumber\\
 & \leq\int_{0}^{T}dt\,\left|\sin\left(\Delta\left(t-T\right)\right)\sin\theta\sin\alpha\right| \leq T \,.
\end{align}
Let's assume $\left|\Delta\right|T<1$ then as before we get
\begin{eqnarray}
Q &\ge& (1 - |\Delta| |\mbox{Im}(N)|)^2
\ge (1- |\Delta| T)^2\,.
\label{bound_Q}
\end{eqnarray}
So $Q>0$ if $\left|\Delta\right|T<1$, i.e. a necessary condition for $Q=0$ is $T\geq\frac{1}{|\Delta|}$.

Using similar arguments to the ones in Sect. \ref{sect_2level_properties}, we see that in this case
as well there can be no $\Delta$-independent scheme with $Q=0$ for all $|\Delta| > 1/T$.
Moreover, an approximation of $Q$ in the case of $|\Delta| T\gg1$ can be
derived in a similar way as in the previously mentioned section.
So we get for $|\Delta| T\gg1$ that
\begin{eqnarray}
Q = \frac{1}{\Delta^2} \dot\alpha (0)^2 + ...
\end{eqnarray}
taking into account the boundary conditions.

%
\begin{figure}
\begin{center}
\includegraphics[angle=0,width=0.8\linewidth]{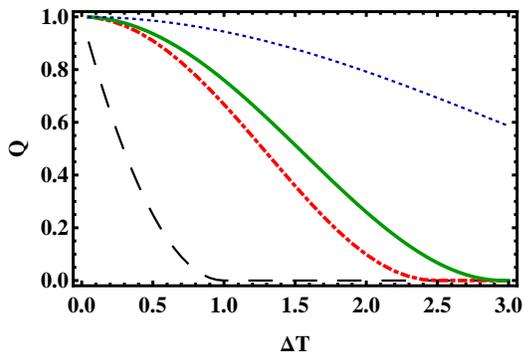}
\end{center}
\caption{\label{fig_6_Q} (Color online) Transition sensitivity $Q$ versus
  $\Delta T$ for different schemes;
reference example($\epsilon=0.002$) from \cite{sta_3level}
(blue, thin, dotted line);
numerical scheme 1 given by Eq. \eqref{scheme_4l_num1} (red, thick, dot-dashed line);
numerical scheme 2 given by Eq. \eqref{scheme_4l_num2} (green,
  solid line);
lower bound for $Q$ as in Eq. \eqref{bound_Q} (black, dashed line).}  
\end{figure}
%

%
\begin{figure}
\begin{center}
(a) \includegraphics[angle=0,width=0.8\linewidth]{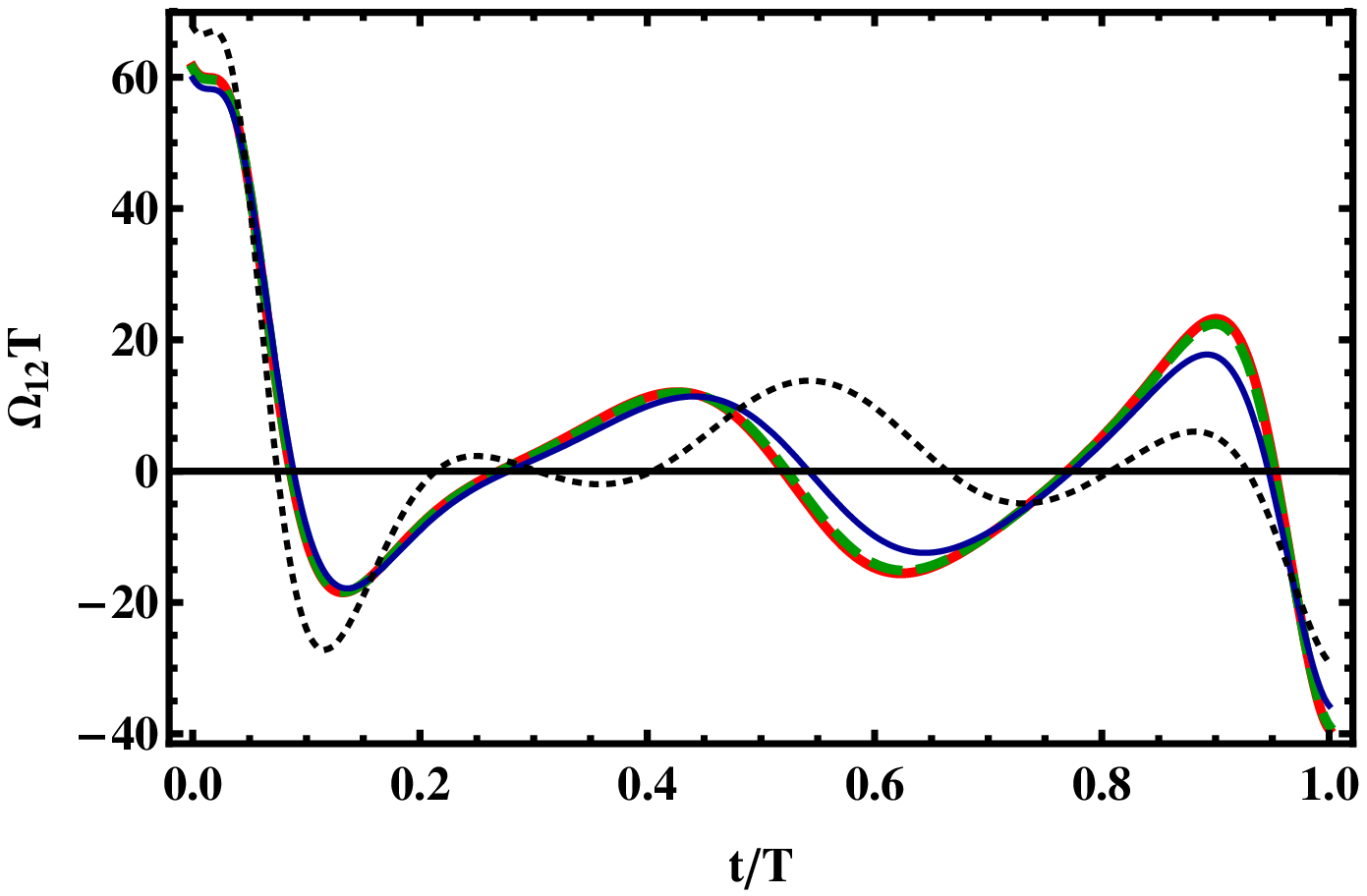}

(b) \includegraphics[angle=0,width=0.8\linewidth]{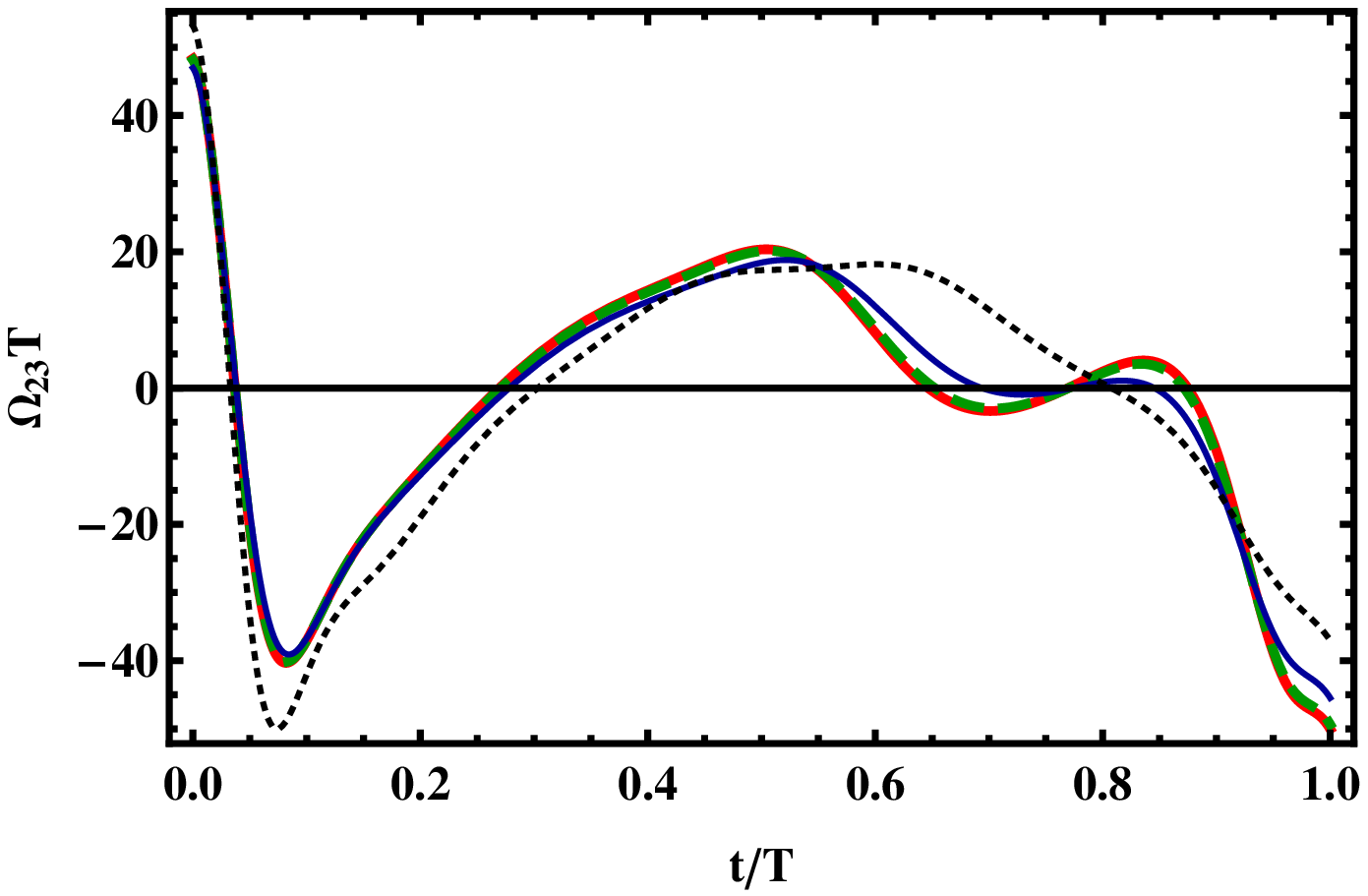}
\end{center}
\caption{\label{fig_7_num1}Rabi frequencies for the numerically optimized
  scheme 1 in Eq. \eqref{scheme_4l_num1} versus time $t$;
(a) Rabi frequency $\Omega_{12}$; (b)Rabi frequency $\Omega_{23}$;
$\Delta T = 0.2$ (red, thick, solid line),
$\Delta T = 1.0$ (green, dashed line),
$\Delta T = 2.0$ (blue, thin, solid line),
$\Delta T = 3.0$ (black, dotted line).}   
\end{figure}
%

%
\begin{figure}
\begin{center}
(a) \includegraphics[angle=0,width=0.8\linewidth]{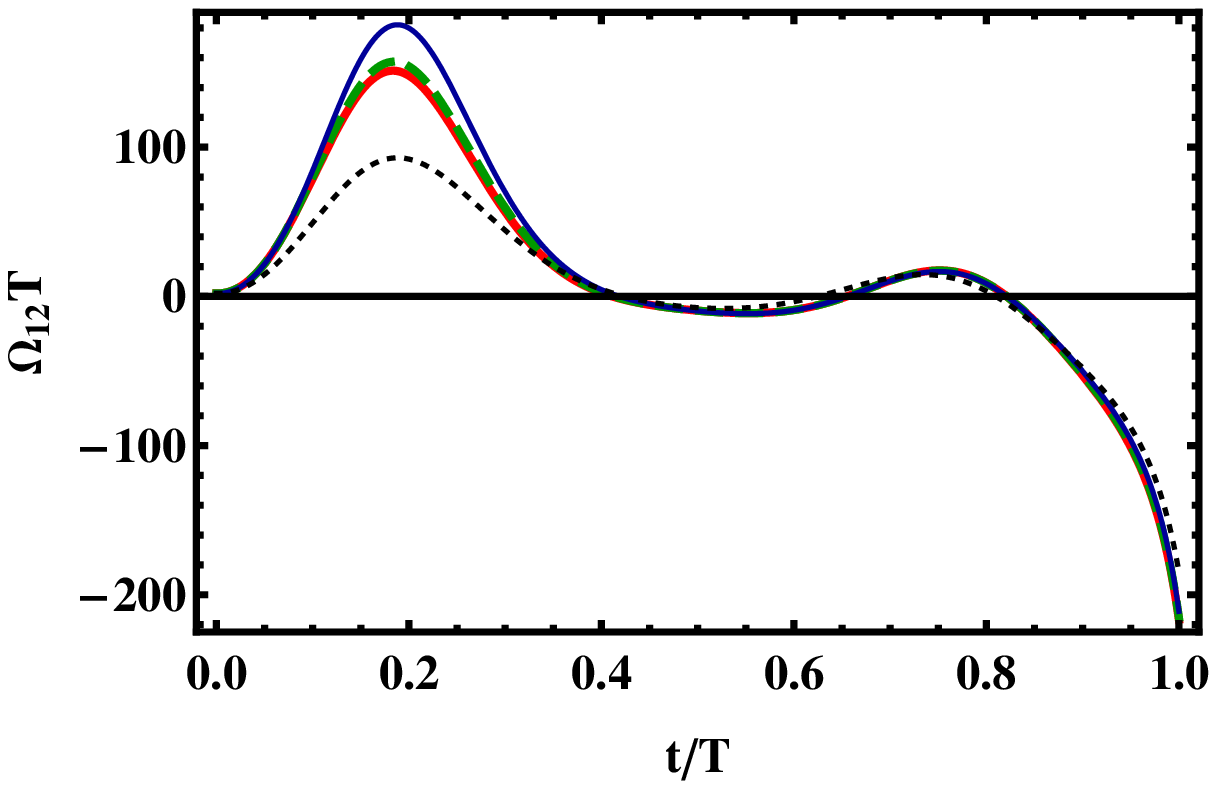}

(b) \includegraphics[angle=0,width=0.8\linewidth]{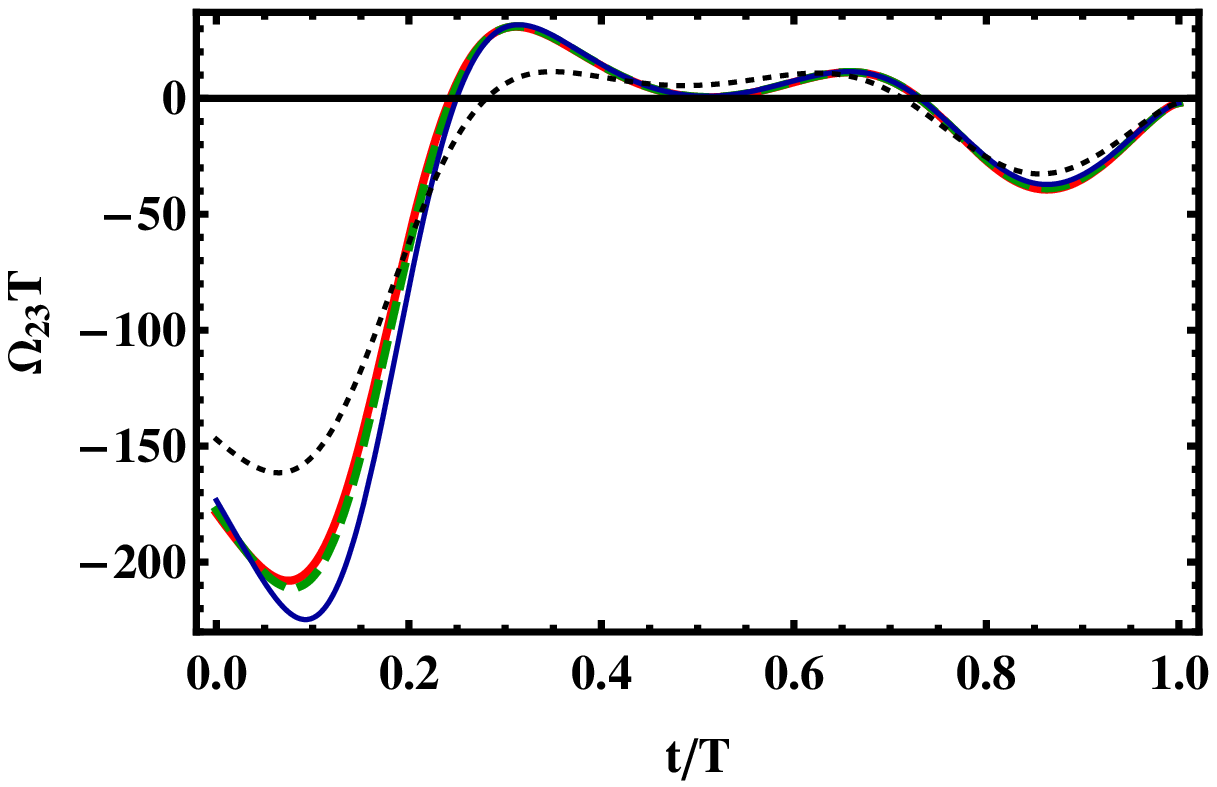}
\end{center}
\caption{\label{fig_8_num2}Rabi frequencies for the numerically optimized
  scheme 2 in Eq. \eqref{scheme_4l_num2} versus time $t$;
(a) Rabi frequency $\Omega_{12}$; (b)Rabi frequency $\Omega_{23}$;
$\Delta T = 0.2$ (red, thick, solid line),
$\Delta T = 1.0$ (green, dashed line),
$\Delta T = 2.0$ (blue, thin, solid line),
$\Delta T = 3.0$ (black, dotted line).}   
\end{figure}
%

\subsection{Example of schemes}

As a reference case we consider one of the protocols given in
\cite{sta_3level}. In this protocol, the following ancillary functions are used 
\begin{eqnarray}
\theta\left(t\right)=\epsilon-\frac{\pi}{2} \, , \, \alpha\left(t\right)=\frac{\pi t}{2T}
\end{eqnarray}
where $0<\epsilon\ll1$ and the only difference
in boundary conditions being that now $\theta\left(T\right)=-\frac{\pi}{2}$.
It should be noted that this protocol does not have perfect population
transfer since the boundary conditions are not exactly fulfilled for
a non-zero $\epsilon$. In \cite{sta_3level} $\epsilon=0.002$ was deemed sufficient. This protocol has the following Rabi frequencies:
\begin{eqnarray}
\Omega_{12}\left(t\right)&=&\frac{\pi}{T}\cot\epsilon\sin\left(\frac{\pi t}{2T}\right)\,,\nonumber\\
\Omega_{23}\left(t\right)&=&\frac{\pi}{T}\cot\epsilon\cos\left(\frac{\pi t}{2T}\right)\,.
\end{eqnarray}
The transition sensitivity for this scheme is shown in Fig. \ref{fig_6_Q}. Here we note that the derivation of the transition sensitivity is based on exact population transfer in the error free case. Hence it is not strictly correct to consider the transition sensitivity for this protocol. However for the purposes of comparison we include it.

In the following we provide two examples of numerically optimized schemes leading to zero transition sensitivity for some range of $\Delta$.
For the first scheme we use the ansatz
\begin{eqnarray}
\theta(t) &=& -\frac{\pi}{2} + (\pi-c_0-c_1) \frac{t}{T} + c_0 \left(\frac{t}{T}\right)^{2} + c_1
\left(\frac{t}{T}\right)^{3}\,,\nonumber\\
\alpha(t) &=& \frac{\pi}{4} \fsin{\theta(t)} + \frac{\pi}{4}
\label{scheme_4l_num1}
\end{eqnarray}
where the parameters $c_0$ and $c_1$ were numerically calculated in order to 
minimize $Q$ for a given $\Delta$. Note that this ansatz
automatically avoids any divergences of the corresponding physical potentials
for $0 \le t \le T$.
The resulting transition sensitivity $Q$ is shown in Fig. \ref{fig_6_Q}.
As it can be seen, 
we can construct schemes which make $Q$ vanish
for $|\Delta| T \ge 2.5$. The corresponding Rabi frequencies
$\Omega_{12}$ and $\Omega_{23}$ are shown in Fig. \ref{fig_7_num1}
for different values of $\Delta T$.

Another example of a scheme is the following
\begin{eqnarray}
\lefteqn{\theta\left(t\right)=-\frac{\pi}{2}-\frac{8(\pi-2d_{0})t^{4}}{T^{4}}+\frac{2t^{3}(-16d_{0}+T+7\pi)}{T^{3}}} & &\nonumber\\
& & - \frac{t^{2}(-16d_{0}+3T+5\pi)}{T^{2}}+t\,,
\nonumber\\
\lefteqn{\alpha\left(t\right)=\frac{1}{2}(2\pi d_{1}+3\pi)\frac{t^{2}}{T^{2}}} & &\nonumber\\
& & +\left(\frac{1}{2}(-2\pi d_{1}-3\pi)+\frac{3\pi}{2}\right)\frac{t}{T}
+ d_{1}\sin\left(\frac{\pi t}{T}\right)-\pi\frac{t^{3}}{T^{3}}\nonumber\\
\label{scheme_4l_num2}
\end{eqnarray}
where the parameters $d_{0}$ and $d_{1}$were numerically calculated
to minimize $Q$ for a given $\Delta$.
$d_{0}$ was restricted to the range $0.55\leq d_{0}\leq2.5$
to avoid divergence of the Rabi frequencies for all $0 \le t \le T$.
The transition sensitivity
$Q$ for this scheme is shown in Fig. \ref{fig_6_Q}. It achieves $Q=0$ at $\Delta T=3$.
The corresponding Rabi frequencies are shown in Fig. \ref{fig_8_num2}.

\subsection{Comparison of the transition probability}
\begin{figure}[t]
\begin{center}
(a) \includegraphics[angle=0,width=0.7\linewidth]{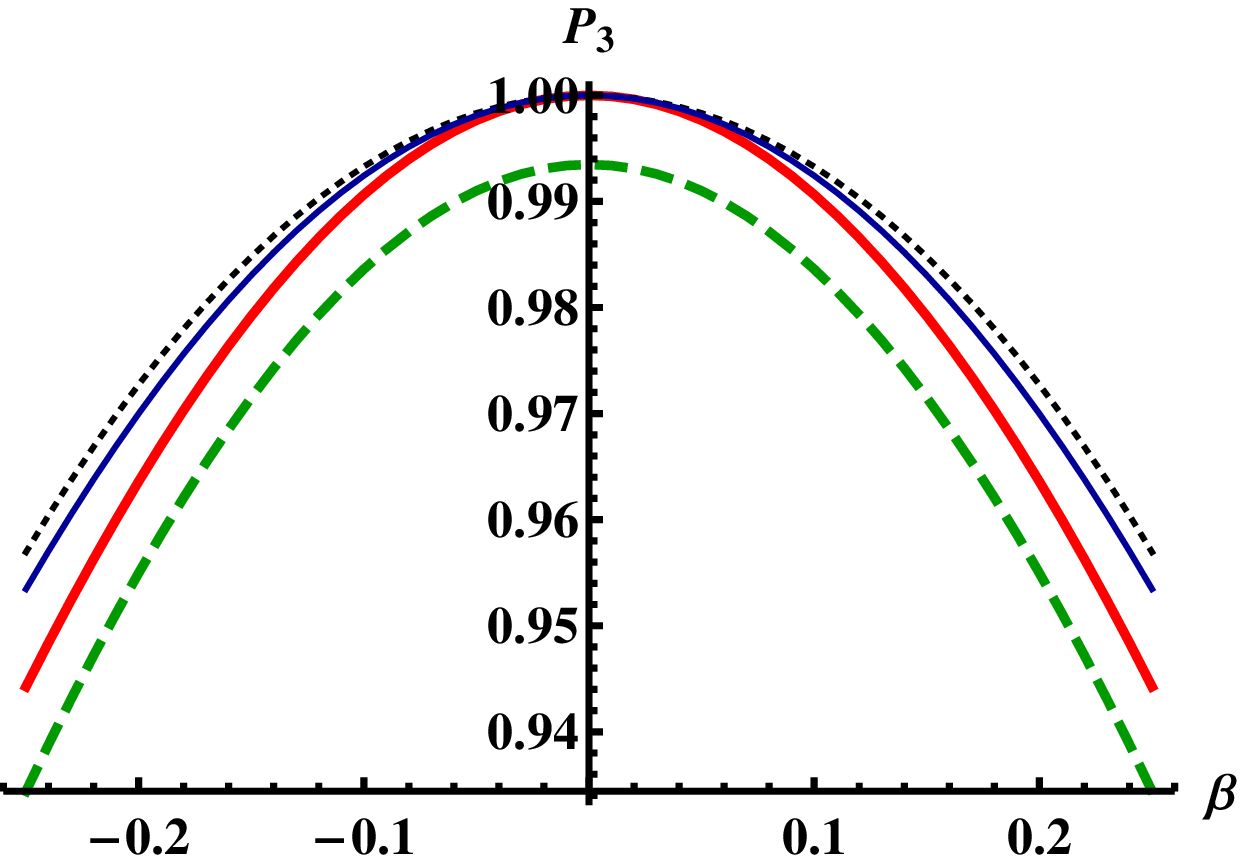}

(b) \includegraphics[angle=0,width=0.7\linewidth]{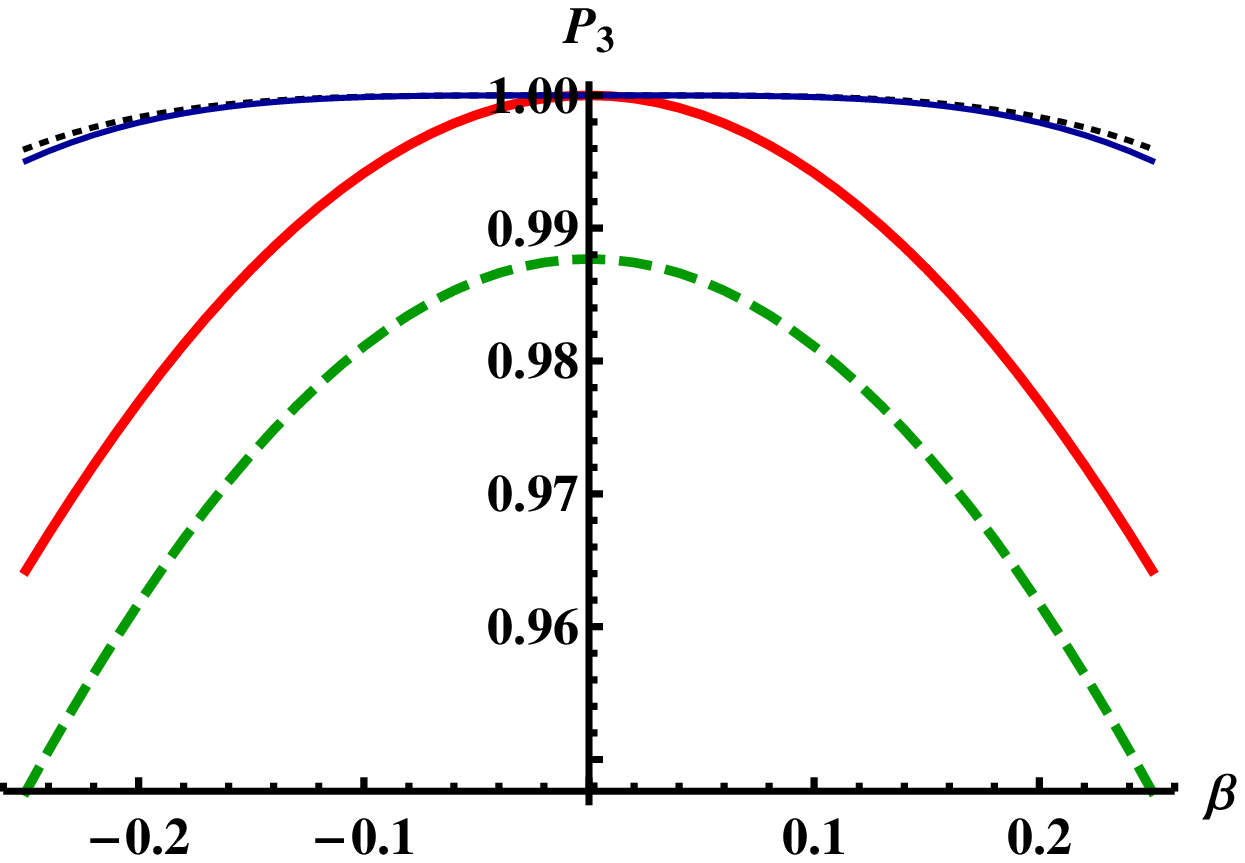}
\end{center}
\caption{\label{fig_9_P3} (Color online) Transition probability $P_3$ versus perturbation strength $\beta$ for different schemes:
reference example($\epsilon=0.002$) from \cite{sta_3level}
(red, thick, solid line);
numerical scheme 1 given by Eq. \eqref{scheme_4l_num1} (black, dotted line);
numerical scheme 2 given by Eq. \eqref{scheme_4l_num2} (blue, thin, solid line);
adiabatic scheme (green, thick, dashed line);
(a) $\Delta T = 1.0$, (b) $\Delta T = 3.0$.}  
\end{figure}

In order to compare the schemes we once again look at the exact (numerically calculated) transition probability $P_{3}$ as a function of $\beta$ as in Fig. \ref{fig_9_P3}.
%
We compare the scheme of the schemes proposed in \cite{sta_3level} as a reference scheme, the numerical scheme 1 given by Eq. \eqref{scheme_4l_num1} and the numerical scheme 2 given by \eqref{scheme_4l_num2}. Once again we see that the transition sensitivity is a good indicator of a stable scheme.
We also consider the area of the pulse and its energy which in this case is defined as $A :=  \int_0^T dt\,
\sqrt{\Omega_{12}^2 + \Omega_{23}^2}$ and $E:=\hbar \int_0^T dt\, \left(\Omega_{12}^2 +
\Omega_{23}^2\right)$ respectively. These values are shown for each scheme in Table \ref{tab_2}.
%
\begin{table}
\begin{tabular}{|c|c|c|}
\hline 
 & $A[\pi]$ & $E[\pi^{2}\hbar/T]$\\
\hline 
Scheme of \cite{sta_3level} $\left(\epsilon=0.002\right)$ 
& $500.00$ & $249999$\tabularnewline
\hline 
Numerical Scheme 1, Eq. \eqref{scheme_4l_num1} & &\\
$\Delta T=1.0 \:(c_{0}=-76.546, c_{1}=49.040)$ & $6.71$ & $70.29$\tabularnewline
$\Delta T=3.0 \:(c_{0}=-76.735, c_{1}=46.054)$ & $6.61$ & $73.61$\tabularnewline
\hline 
Numerical Scheme 2, Eq. \eqref{scheme_4l_num2} & &\\
$\Delta T=1.0 \:(d_{0}=0.794, d_{1}=-15.633)$ & $24.34$ & $1171.7$\tabularnewline
$\Delta T=3.0 \:(d_{0}=0.852, d_{1}=-13.204)$ & $18.65$ & $663.17$\tabularnewline
\hline 
Adiabatic Scheme, Eq. \eqref{pot_stirap} & $\Omega_{0}T\pi^{-1}$ & $\Omega_{0}^{2}T^{2}\pi^{-2}$\\
$\Delta T=1.0$ & $8.38$ & $70.29$\tabularnewline
$\Delta T=3.0$ & $8.58$ & $73.61$\tabularnewline
\hline 
\end{tabular}\caption{Pulse area $A$ and energy $E$ for different protocols.\label{tab_2}}
\end{table}
%

For completeness we also include the following adiabatic STIRAP-like scheme in
our comparison \cite{stirap}:
\begin{eqnarray}
\Omega_{12}&=&\Omega_{0}\sin\left(\frac{\pi t}{2T}\right)\,,\\
\Omega_{23}&=&\Omega_{0}\cos\left(\frac{\pi t}{2T}\right)\,.
\label{pot_stirap}
\end{eqnarray}
$\Omega_{0}$ was chosen so that the adiabatic scheme has the same energy as the numerical scheme 1. 

Both numerically-optimized schemes result in the largest $P_3$ in Fig. \ref{fig_9_P3}(a) if $\beta \neq 0$ for $\Delta T = 1.0$. If $\Delta T = 3.0$, see Fig. \ref{fig_9_P3}(b), then both numerical-optimized schemes result in nearly full population transfer even in the case of $-0.1 < \beta < 0.1$. It can be seen that a full population transfer is not achieved in both cases by this adiabatic scheme for $\beta=0$.

\section{Discussion}

In this paper, we have examined the stability of shortcuts to adiabatic
population transfer in two- and three-level quantum systems against unwanted transitions. For the two-level case as well as for the three-level case, we have defined a
transition sensitivity which quantifies how sensitive a given scheme is
concerning these unwanted couplings to another level.

We have compared the transition sensitivity of
different schemes in both settings. We also have provided examples
of shortcut schemes leading to a zero transition sensitivity in certain regimes i.e. almost full population inversion is achieved in the presence of unwanted transitions.

This approach could be even further generalized; one could construct different shortcut schemes fulfilling even further constraints apart from vanishing
transition sensitivity similar to \cite{Andreas2013}.
This work could also be generalized to different level structures of the unwanted transitions or to multiple unwanted transition channels.
In the latter case, one might expect to find that the unwanted transition with lowest detuning would dominate.

\section*{Acknowledgments}

We are grateful to David Rea for useful discussion and commenting on the manuscript.


\end{document}